\documentclass[12pt,preprint]{aastex}
\usepackage{amssymb}
\newcommand{\fracdps}[2]{\frac{\displaystyle #1}{\displaystyle #2}}
\newcommand{\dlagr}[1]{\fracdps{d #1}{dt}}

\newcommand{\Ll}{\mathcal{L}}
\newcommand{\kboltz}{{\cal R}}
\newcommand{\kbf}{{\bf k}}
\newcommand{\rbf}{{\bf r}}
\newcommand{\xbf}{{\bf x}}
\newcommand{\vbf}{{\bf v}}
\newcommand{\tchi}{{\tau_\chi^{}}}
\newcommand{\vperp}{{\vbf_1}_\perp^{}}

\newcommand{\defdef}{{\buildrel \rm def \over =}}
\newcommand{\Ss}{\psi}

\newcommand{\tprim}{{t^{\prime\,}}}
\newcommand{\funcool}{\Lambda}
\newcommand{\LambdaCoulomb}{\Lambda_{C}}
\newcommand{\Omgc}{\Omega_{dc}}
\newcommand{\kcrit}{k_{crit}}
\newcommand{\lambcrit}{{\lambda_{crit}}}

\newcommand{\Omc}{\Omega_{c}}
\newcommand{\Omcdrho}{\Omega_{c\rho}}
\newcommand{\Omcdp}{\Omega_{cp}}
\newcommand{\OmcdrhoF}{{\Omega_F}}
\newcommand{\ktilde}{\tilde{k}}
\newcommand{\tauchiJ}{{{\tau_{\chi}^{}}_J^{}}}

\shorttitle{THERMAL INSTABILITY IN A COOLING MEDIUM}
\shortauthors{GOMEZ-PELAEZ & MORENO-INSERTIS}

\begin{document}

\slugcomment{Accepted by ApJ. Probable publication date: April 20, 2002 }

\title{THERMAL INSTABILITY IN A COOLING AND EXPANDING MEDIUM
INCLUDING SELF-GRAVITY AND CONDUCTION}

\author{A.J. GOMEZ-PELAEZ\altaffilmark{1} AND F. MORENO-INSERTIS\altaffilmark{1,2}}
\affil{\altaffilmark{1} Instituto de Astrof\'{\i}sica de Canarias, E-38200 La Laguna
(Tenerife), Spain}

\affil{\altaffilmark{2} Department of Astrophysics, University of La Laguna,
E-38200 La Laguna (Tenerife), Spain}
\email{ajgomez@ll.iac.es and fmi@ll.iac.es}

\begin{abstract}
A systematic  study of the  linear thermal stability  of a medium  subject to
cooling, self--gravity  and thermal  conduction is carried  out for  the case
when  the unperturbed state  is subject  to global  cooling and  expansion. A
general,  recursive WKB  solution for  the perturbation  problem  is obtained
which can  be applied to a  large variety of  situations in which there  is a
separation of  time-scales for  the different physical  processes.  Solutions
are explicitly given and discussed for the case when sound propagation and/or
self--gravity are the fastest  processes, with cooling, expansion and thermal
conduction operating on slower time-scales.  A brief discussion is also added
for the solutions in the cases  in which cooling or conduction operate on the
fastest time-scale.
The general WKB  solution obtained in this paper  permits solving the problem
of the effect of thermal conduction and self-gravity on the thermal stability
of a globally cooling and expanding  medium. As a result of the analysis, the
critical  wavelength (often called  {\it Field  length}) above  which cooling
makes the perturbations unstable against  the action of thermal conduction is
generalized to the case of an  unperturbed background with net cooling. As an
astrophysical application,  the {\it generalized Field  length} is calculated
for  a  hot ($10^4  -  10^8$  K), optically  thin  medium  (as pertains,  for
instance, for  the hot interstellar medium  of SNRs or  superbubbles) using a
realistic  cooling  function  and  including  a  weak  magnetic  field.   The
stability  domains are compared  with the  predictions made  on the  basis of
models for which  the background is in thermal  equilibrium.  The instability
domain of the sound waves, in particular, is seen to be much larger in the
case with net global cooling.

\end{abstract}


\section{INTRODUCTION}

Thermal instability often appears in astrophysical media which are undergoing
a global process of cooling, expansion or contraction.  A non-exhaustive list of
examples includes: thermal instability in a cooling flow in a galaxy cluster
\citep{mathews,balbussoker89,balbus91}, in a stellar wind \citep{balbussoker89}, 
or in the hot ISM \citep{begelman}, and structure formation in a collapsing protogalaxy
\citep{fallrees85}. In all those cases, the unperturbed state is not
one of thermal or dynamical equilibrium, so that the stability and the time
evolution of the perturbations cannot be studied by means of a classical
normal--mode Fourier analysis in time. The situation is complicated by other
physical processes which can be important for the unperturbed state or for
the perturbation itself, like self-gravity, thermal conduction, and magnetic
induction and diffusion.  The stability criteria and the evolution of the
perturbations directly depend on the relative time-scale of the latter
processes with respect to each other and to the time-scales of cooling and
expansion of the background in which they are developing.

The astrophysical literature abounds in examples of thermal stability
analyses and their application to specific problems, as in \citet{parker},
\citet{weymann}, \citet{field65}, \citet{gold}, \citet{defouw},
\citet{heyvaerts}, \citet{mathews}, \citet{flan}, \citet{balbus86,balbus91},
\citet{malagoli}, \citet{balbussoker89}, \citet{begelman}, \citet{loew},
\citet{chun-rosner}, \citet{burkert}. A milestone was the paper by
\citet{field65}. He considered a non-expanding homogeneous medium in thermal
equilibrium (i.e., with cooling exactly compensating heating), including
thermal conduction and magnetic fields. The small--perturbation problem is
then amenable to a normal-mode Fourier analysis in space and
time. Additionally to magnetosonic waves, which can be strongly modified by
the thermal processes, he found a non-oscillatory mode (called condensation
mode) which may have an isobaric or isochoric character depending on the
amount of cooling or conduction possible during one crossing time of a
magnetosonic wave across the perturbation. Conduction becomes important at
small spatial scales and, in general, has a strongly stabilizing
effect. Another important step forward was taken by \citet{mathews} and
\citet{balbus86}, who considered a time--dependent unperturbed medium.  By
using the entropy equation, they obtained stability criteria for the
condensation mode in a medium with global cooling, not taking into account
conduction, self-gravity or any background stratification. \citet{balbus86},
in particular, carried out a WKB analysis in space for radial perturbations
in an expanding or contracting medium with spherical symmetry.
\citet{balbussoker89} relaxed the condition of homogeneous background and
obtained solutions for a stratified medium. \citet{balbus91} also included a
background magnetic field.

The topic of thermal instability in astrophysical media seems to have been
gathering momentum in the past several years. In spite of the foregoing,
there is to date no {\it systematic} study of the stability of a magnetized
medium subject to cooling, self-gravity and thermal conduction when the
unperturbed state itself is undergoing expansion and cooling. The purpose of
the present paper is to carry out such a study using, wherever
possible, a WKB expansion {\it in time}. The treatment introduced here
permits recovering many results of the previous literature in a unified
manner and allows obtaining new ones. The non-magnetic and non-stratified
case is considered in the present paper, which deals with the problem of a
uniform (but time-dependent) background. However, the results of this
paper can also be applied locally to a weakly non-homogeneous background (see
sect. \ref{swkb}). Several characteristic time-scales have to be considered,
to wit, those associated with the background expansion, self-gravity,
cooling, sound-wave propagation and conduction.  When background cooling and 
expansion are {\it not} the fastest processes in the system, a WKB analysis in
time can be carried out. General WKB solutions up to arbitrary order are then
obtained and applied to the particular cases when the fastest time-scales are 
those associated with sound wave propagation, with sound and self-gravity simultaneously, 
or with conduction.

New results obtained in the present paper are as follows.  First, we obtain
an expression for the critical length separating stable from unstable domains
which generalizes the classical {\it Field length} to a system which, in the
unperturbed state, is cooling (or otherwise) and expanding or contracting globally.
\citet{field65} showed the opposite effects of conduction and cooling on the
stability of a system in static and thermal equilibrium in the case when the
cooling function, $\Ll$, fulfills $[\partial \Ll/\partial T]_p < 0 $, and
found that below a critical wavelength (the classical {\it Field length})
conduction stabilizes the system. Our study permits relaxing the condition
of thermal equilibrium in the unperturbed state and including in the latter a
global expansion or contraction. We may note that, for lack of a more
accurate expression, different authors in the past have tentatively used the
value calculated by Field even though their problem involved a medium
undergoing global cooling \citep[e.g.,][]{david89,pistinner}.
In section \ref{s_astr}, we calculate the generalized Field length for a
medium undergoing a realistic net cooling, and compare it with the classical Field length of
a medium in thermal equilibrium with the same cooling but with a balancing constant heating. The
results of the present paper also generalize those of \citet{balbus86}, who
considered a medium undergoing global cooling in the unperturbed state, but
ignored thermal conduction.

Second, there is in the literature no simultaneous study of the thermal and
gravitational linear stability of a medium undergoing net cooling. The combination
of both phenomena is important, for example, for the structure formation in a
protogalaxy \citep{fallrees85}, where condensations previously formed by
thermal instability in a cooling medium become gravitationally unstable. In
section \ref{s_sougr} we calculate how the acoustic-gravity modes of
\citet{jeans} are modified through conduction and cooling when the background
itself is cooling globally. We also show the existence of a hitherto unknown
condensation mode in this case. In all these solutions the perturbations can
go through different stability regimes: in a cooling medium the
generalized Field length and the Jeans length change with time whereas the
wavelength of the perturbation remains constant.

There is an alternative to a WKB expansion when trying to solve
the problem tackled in this paper. \citet{burkert} have used a Taylor
expansion in time for the reduced problem when there is no thermal
conduction, self-gravity or background expansion.  Yet, a Taylor expansion
has the large disadvantage of being valid only at times small compared with
the characteristic times of evolution of the background. This is in contrast
to the WKB approach taken in the present paper, which is valid for arbitrary
times as long as the perturbation does not grow beyond the linear regime.

The present paper is organized as follows: in section \ref{s_pose_problem},
the equations of the problem are presented and the relevant time-scales
introduced both for the background and the perturbations.  Section \ref{swkb}
contains a discussion of the application of the WKB method to the present
problem and provides a general WKB solution to the linear perturbation
equations. In section \ref{sec_intermediate}, a collection of solutions for
intermediate wavelengths is presented. The word {\it intermediate} is used in
the sense that the wavelength is not small nor large enough for conduction or
cooling, respectively, to be the dominant processes in the evolution.  The
case in which sound propagation is the fastest process is considered in
section \ref{ssou} whereas in section \ref{s_sougr} both self-gravity and
sound are the fastest processes. Section \ref{sother} is devoted to
the large and small wavelength regimes and, in particular, to the cases when cooling or
conduction operate on the fastest time-scale. In section \ref{s_astr}, the
generalized Field length and the instability domain are computed for an
optically thin medium at high temperature threaded by a magnetic field
without dynamical effects (e.g., the hot ISM). The
final section (sect.~\ref{ssum}) contains a summary and further discussion of
the results of the paper.

\section{THE EQUATIONS OF THE PROBLEM AND THEIR WKB SOLUTION}\label{s_pose_problem}

This section discusses the physical processes relevant for the present
problem, the governing system of equations, and the method used to solve this
system. The equations are given first in general (sec.~\ref{sequ}) and then 
specialized for the homogeneous unperturbed medium (sec.~\ref{sback}) and for
small perturbations to that medium (sec.~\ref{slinear}). There is a considerable number of different
physical processes of importance for the present problem: the corresponding
time-scales are listed and explained in section \ref{stimes}. A general WKB
solution to the linear perturbation system is obtained and discussed in section \ref{swkb}.

\subsection{Equations}
\label{sequ}

We consider an optically thin polytropic ideal gas. The fluid equations that describe
the evolution of the system including thermal conduction
and self-gravity are:

\begin{equation} \label{eq_cont}
\dlagr{\rho}+\rho {\bf \nabla \cdot v} =0 \; ,
\end{equation}

\begin{equation} \label{eq_mom}
\rho \dlagr{{\bf v}}+ {\bf \nabla} p + \rho {\bf \nabla} \Phi =0 \; ,
\end{equation}

\begin{equation}\label{eq_ener}
\frac{\kboltz}{\mu (\gamma -1)}\rho \dlagr{T}  + p {\bf \nabla \cdot v} +\rho\Ll
-{\bf \nabla \cdot}( \chi {\bf \nabla} T)=0 \; ,
\end{equation}
where $d/dt=\partial/\partial t + {\bf v \cdot \nabla}$ is the Lagrangian time
derivative, $\gamma$ is the polytropic index of the ideal gas, $\mu$ is the
mean atomic mass per particle, $\kboltz$ is the universal gas constant,
$\Ll(\rho,T)$ (erg s$^{-1}$ gr$^{-1}$) is the net cooling function (including
heating), $\chi(\rho,T)$ is the thermal conduction coefficient and $\Phi$
is the gravitational potential caused by the mass distribution of the system
$\Delta \Phi = 4 \pi G \rho $, with $G$ the universal
constant of gravitation. All other symbols have their usual meaning in
hydrodynamics.  Equations (\ref{eq_cont}), (\ref{eq_mom}) and (\ref{eq_ener}) are
the continuity, momentum and energy equations, respectively. The temperature
is related to the other quantities by the ideal gas equation:
$p=(\kboltz/\mu)\rho T$. 
The change of $\mu$ due to recombination or ionization processes is not
considered in this paper. This is fully acceptable for high enough temperature
($T>10^5$K), in which the medium is almost completely ionized, or in
situations at lower temperature when the change in $\mu$ is less important
than the contribution of cooling and conduction. The thermal instability of a
medium in thermal equilibrium with variable $\mu$ has been considered by \citet{gold} and
\citet{defouw}. Finally, note that the radiative flux in the diffusion limit 
(optically thick medium) is identical to the classical thermal conduction
flux except for the associated conductivity coefficient
\citep[e.g.,][]{mihalas}. Therefore, our treatment also allows for the use of
radiative transport in the diffusion limit.

\subsection{Background Evolution}
\label{sback}

As a basis for the small--perturbation analysis, we assume a homogeneous 
background which is expanding uniformly and cooling (or heating) by
radiation, so that the unperturbed quantities only depend on time. The
background quantities will be denoted with the subscript $0$. The expansion
is given by
\begin{equation} \label{eq_expan}
\rbf = a(t)\,\xbf \; ,
\end{equation}
where $\rbf$ is the Eulerian coordinate, $\xbf$ is the Lagrangian
coordinate and $a(t)$ is the expansion parameter. Using equation (\ref{eq_expan}), 
the unperturbed velocity field is given by 
\begin{equation} \label{eq_v0}
\vbf_0^{}(\rbf, t)=\frac{\dot{a}}{a}\,\rbf \; ,
\end{equation}
where the dot over a symbol (as in $\dot{a}$) indicates its time derivative.

Equations (\ref{eq_cont}) and (\ref{eq_ener}) reduce for
the background to
\begin{eqnarray}
\noalign{\vspace{2mm}}
\dlagr{\rho_0} &=& -3 \frac{\dot{a}}{a}\rho_0 \; , \label{eq_cont0}\\
\noalign{\vspace{2mm}}
\frac{d}{dt}\log{T_0} &=& -(\gamma-1)\left(3 \frac{\dot{a}}{a}+\frac{\mu}{\kboltz}
        \frac{\Ll_0}{T_0}\right) \; ,\label{eq_ener0}\\
\nonumber
\end{eqnarray}
respectively, where $\Ll_0=\Ll[\rho_0(t),T_0(t)]$. The time dependence of
$a(t)$ must be specified from equation~(\ref{eq_mom}). The latter
yields
\begin{equation} \label{eq_mom00}
\dot{a} = {\rm const} 
\end{equation}
when self-gravity is negligible and
\begin{equation} \label{eq_mom0}
\ddot{a} = -\frac{4}{3}\pi G \rho_0 a \; 
\end{equation}
 otherwise. 
Additionally,  more general forms for $a(t)$ are allowed for the local application of
the perturbative analysis around a Lagrangian fluid element in a non-uniform
background, as will be explained at the end of section \ref{swkb}. 

\subsection{Linearized Equations for the Perturbations}
\label{slinear}

To obtain a linearized system of equations for the perturbations we proceed
as follows. First, we split each variable into unperturbed and perturbed
components, indicating the latter with a subscript $1$, like, e.g.
\begin{eqnarray}
\rho_1^{}(\rbf,t) &\defdef& \rho(\rbf,t)-\rho_0^{}(t) \label{eq_pertdens}\,,\\
\vbf_1^{}(\rbf,t) &\defdef& \vbf(\rbf,t)-\vbf_0^{}(\rbf,t) \label{eq_pertvel}\,,
\end{eqnarray}
and similarly for pressure and temperature. Eulerian linear perturbations are
then taken in equations (\ref{eq_cont}) through (\ref{eq_ener}). Thereafter, the
Eulerian divergence operator is applied to the momentum equation and all
equations are rewritten in terms of the Lagrangian coordinate $\xbf$ defined in
equation (\ref{eq_expan}). The resulting linear system has coefficients which depend
on $t$ but not on $\xbf$. We then carry out a spatial Fourier analysis, with
Fourier components proportional to $\exp(i\,{\bf k \cdot x})$, so that $\kbf$
is the Lagrangian wave vector:
\begin{eqnarray}
{\rho_1}_\kbf^{}(\xbf,t)   &=& {\rho_1}_\kbf^{}(t)\,\exp(i\,{\bf k \cdot x})
\,,\label{eq_densfourier}\\
{\vbf_1}_\kbf^{}(\xbf,t) &=& {\vbf_1}_\kbf^{}(t)\,\exp(i\,{\bf k \cdot x}) 
\,,\label{eq_velfourier}\,
\end{eqnarray}
and expressions similar to (\ref{eq_densfourier}) for all other scalar
variables (pressure, temperature, etc). For simplicity, we eliminate the
index $\kbf$ from the Fourier amplitudes in the following. The perturbed
velocity is then split into its components parallel and normal to $\kbf$: 
\begin{equation}
 \vbf_1^{}(t) = v_1^{}(t) \frac{\kbf}{k} + \vperp (t)\,. 
\end{equation}
Finally, the resulting linear system is simplified by repeated use of the background
equations and elementary thermodynamic relations. We obtain:
\begin{eqnarray}
\noalign{\vspace{2mm}}
\frac{d}{dt}\left(\frac{\rho_1}{\rho_0}\right) &=& -i \ktilde\, v_1 \; ,\label{eq_rho1}\\
\noalign{\vspace{2mm}}
\dlagr{v_1} &=& -\frac{\dot a}{a}\, v_1 -i \frac{\omega_s c_s}{\gamma} \,\frac{p_1}{p_0} +
i \fracdps{\omega_g^2}{\ktilde} \, \frac{\rho_1}{\rho_0} \; ,\label{eq_v1}\\
\noalign{\vspace{2mm}}
\frac{d}{dt}\left(\frac{p_1}{p_0}\right) &=& -i \gamma \ktilde \, v_1 -
\left(\Omcdrho + \tchi^{-1}\right)  \frac{p_1}{p_0} 
+ \left(\Omcdp + \tchi^{-1}\right) \frac{\rho_1}{\rho_0} \; ,\label{eq_p1}
\end{eqnarray}
with all coefficients of the perturbations being evaluated for the background values.
As a result of the linearization, the time derivative $d/dt$ in
equations (\ref{eq_rho1}) through (\ref{eq_p1}) is now given by $d/dt
=\partial/\partial t + \vbf_0 \bf{\cdot \nabla}$, i.e., it is the material
derivative {\it as one moves with the unperturbed fluid}.  The other symbols
have the following value:
\begin{equation}\label{eq_coe2}
\omega_s = \ktilde c_s \;; \qquad \omega_g^2 = 4 \pi G \rho_0 \;;
\end{equation}
\begin{equation}\label{eq_coe3}
        \tchi^{-1} = (\gamma-1)\frac{\mu}{\kboltz}\frac{\chi_0}{\rho_0}\ktilde^2\;;
\end{equation}
\begin{eqnarray}\label{eq_coe4}
\Omcdrho  = (\gamma-1)\frac{\mu}{\kboltz} T_0 \left[\frac{\partial}{\partial T}
\left(\frac{\Ll}{T}\right) \right]_\rho \qquad \hbox{and}\qquad
\Omcdp  = (\gamma-1)\frac{\mu}{\kboltz} T_0 \left[\frac{\partial}{\partial T}
\left(\frac{\Ll}{T}\right) \right]_p \;,\qquad 
\end{eqnarray}
\noindent where 
\begin{equation}\label{eq_coe5}
c_s =\left( \gamma \fracdps{p_0}{\rho_0} \right)^{1/2} \qquad\hbox{and}\qquad
\ktilde = \fracdps{k}{a} 
\end{equation}

\noindent are, respectively, the adiabatic background sound speed
and the Eulerian wavenumber. The coefficients
(\ref{eq_coe2})--(\ref{eq_coe4}) provide  characteristic 
times for our problem: their meaning and relevance will be discussed in the next section.
One also obtains an independent equation for the perpendicular component
of the velocity, viz.:
\begin{equation}\label{eq_vperp}
\dlagr{}\nabla \times (a^2\, \vbf_1)=0 \;,
\end{equation}
i.e., the integral of the perturbed velocity along a closed material curve remains
constant in time. 

The system (\ref{eq_rho1})--(\ref{eq_p1}) is a linear ODE system in $t$ for the
vector of unknowns $(\rho_1/\rho_0, \, v_1,\, p_1/p_0)$. Considering 
the ratio perturbation/background is  the natural choice because the
unperturbed quantities depend on time. {\it We will say that there is
instability if either $|\rho_1/\rho_0|$ or $|T_1/T_0|$ grow with time.}
The velocity perturbation $\vbf_1$ is not compared with any
unperturbed quantity because there is no single unperturbed velocity which
could serve as reference at all length scales: at scales smaller than the Jeans length, 
the sound velocity should be used whereas one should choose  $\sqrt{4 \pi G \rho_0}/k$
for larger scales.

\subsection{Characteristic Timescales}
\label{stimes}

A fundamental timescale in the problem is {\it the period of a sound wave}, $\tau_s
= 2\pi/\omega_s$, with $\omega_s$ defined in equation (\ref{eq_coe2}). Other 
important timescales are as follows:

\subsubsection{Timescales Associated with Background Expansion and with
Self-Gravity}\label{s_tscl_exp_selfgrav}

\noindent (a) {\it The characteristic time of background expansion} is:
\begin{equation}\label{eq_te}
        \tau_{e}=\frac{a}{|\dot{a}|} \;.
\end{equation} 
This time is also the time-scale for the adiabatic temperature decrease
which follows from the expansion (see eq.~[\ref{eq_ener0}]) and for the
inertial change of the perturbed velocity associated with the background
expansion (first term on the right-hand side of eq.~[\ref{eq_v1}]).

\noindent (b) The characteristic {\it growth time of a perturbation by
self-gravity}, $\tau_g$, is
\begin{equation}\label{eq_taug}
\tau_g = \left(4 \pi G \rho_0\right)^{-1/2} \;,
\end{equation}
\noindent which coincides with  $\omega_g^{-1}$, as defined 
in equation (\ref{eq_coe2}). From equation (\ref{eq_mom0}), $\tau_g$ is 
also the {\it timescale for the deceleration of the background expansion}.

\subsubsection{Timescales Associated with Cooling and Conduction}\label{s_tscl_cool_cond}

\noindent (c) The characteristic {\it timescale of background cooling} because of radiation, $\tau_c$,
is given by (see eq.~[\ref{eq_ener0}]) 
\begin{equation}\label{eq_tc}
        \tau_c= \left|\Omc\right|^{-1} \quad \hbox{with} \quad \Omc = (\gamma
-1) \frac{\mu}{\kboltz}\frac{\Ll_0}{T_0} \;.
\end{equation}

\noindent For the growth (or otherwise) of the perturbations, one has to consider not
only the cooling function itself at a given value of $T_0$ and $\rho_0$, but also the {\it differential cooling}
at neighboring values of temperature and density. The precise expression of this
fact is obtained through the perturbed energy equation (eq.~[\ref{eq_p1}]) rewritten in
the form: 
\begin{equation}\label{eq_T1}
\frac{d}{dt}\left(\frac{T_1}{T_0}\right) = - i (\gamma-1) \ktilde \, v_1
- \left(\Omgc - \Omc + \tchi^{-1}\right) \,\frac{T_1}{T_0}\;,
\end{equation} 

\noindent with {\it the growth rate due to differential cooling}, $\Omgc$, defined as
\begin{equation}\label{eq_omgc}
\Omgc = (\gamma -1) \fracdps{\mu}{\kboltz} \left[\left(\fracdps{\partial \Ll}
{\partial \rho}\right)_T \fracdps{\rho_1}{T_1}  +
\left(\fracdps{\partial \Ll}{\partial T}\right)_\rho \right] \;. 
\end{equation}
In the simple case when the perturbation is carried out keeping pressure,
density or entropy constant, $\rho_1/T_1$ in equation (\ref{eq_omgc}) only
depends on the background variables. Then, $\Omgc^{-1}$ and $\tau_c$ are 
the {\it characteristic timescales of cooling}. 

\noindent (d) From equation (\ref{eq_T1}), {\it the timescale of thermal conduction}
is $\tchi$, which was defined in equation (\ref{eq_coe3}).

\vspace{5mm}

The different characteristic times have different dependences on $k$, so the
corresponding processes will become important in different wavelength
domains. For instance, the conduction growth-rate, $\tchi^{-1}$, depends
on $k$ as $k^2$. Hence, conduction dominates at short wavelengths. The
cooling, background expansion and self-gravity time-scales have no spatial dependence.
Hence, these processes are dominant at long wavelengths. The frequency of
sound, finally, depends on the first power of $k$. Hence sound propagation
could be the dominant effect at intermediate wavelengths. The boundaries
between the different regions can be easily estimated using the definitions
of the characteristic times just given.

\subsection{Solution Method: WKB}
\label{swkb}

To find the time evolution of the perturbations, the linear ODE system
(\ref{eq_rho1})-(\ref{eq_p1}) has to be solved.  Unfortunately, no simple
analytical solution can be found in a general case. The only exception is
when all coefficients are time-independent: then, a Fourier analysis in time
can be carried out and the system reduces to an algebraic eigenvalue
problem, from which the normal modes of the problem can be obtained. This
method was used in many of the early papers on thermal instability, like
\citet{field65}. If the background is expanding, or is undergoing net
cooling, though, the coefficients are time-dependent and the system must be solved
numerically in general. However, a general statement about the stability of the
physical system can only be given via analytical solutions, even if they are
only approximations to the actual solutions. In the following, we study the
WKB solutions to the problem. 

To carry out the WKB analysis, we first transform the  system
(\ref{eq_rho1})-(\ref{eq_p1}) into a third order linear ODE 
for $\rho_1/\rho_0$.
\begin{equation}\label{eq_rho3}
\frac{d^3}{d t^3}\left(\frac{\rho_1}{\rho_0}\right)
+ C(t)\, \frac{d^2}{dt^2} \left(\frac{\rho_1}{\rho_0}\right)
+ B(t)\, \frac{d}{d t}\left(\frac{\rho_1}{\rho_0}\right)+ A(t)\, \frac{\rho_1}{\rho_0}=0\;,
\end{equation}
where 
\begin{eqnarray}
\noalign{\vspace{3mm}}
A(t) &=& \frac{\omega_s^2}{\gamma}\left(\Omcdp + \tchi^{-1}\right)+\omega_g^2\left[(4-3\gamma)
\frac{\dot a}{a}-\Omc - \Omcdrho - \tchi^{-1}  \right]\;,\label{eq_cA}\\
\noalign{\vspace{3mm}}
B(t) &=& \omega_s^2-\omega_g^2+2\frac{\ddot{a}}{a} - 2\frac{\dot a}{a} \left[(2-3\gamma)
\frac{\dot a}{a}- \Omc  - \Omcdrho - \tchi^{-1} \right]\;,\label{eq_cB}\\
\noalign{\vspace{3mm}}
C(t) &=& \Omc + \Omcdrho + \tchi^{-1} + (3\gamma +1)\frac{\dot a}{a}\;.\label{eq_cC}
\end{eqnarray}

The WKB analysis requires a clear separation of time scales, in the sense that there is
a group of processes, which we will call {\it dominant}, acting on a
time-scale which is much shorter than the shortest time-scale of the remaining processes
(henceforth called {\it non-dominant} processes).  In addition, for WKB to be
applicable, the time-scales of evolution of the background (net cooling,
expansion) must be in the non-dominant group. Call $\tau_0(t)$ the shortest
of the characteristic times associated with the dominant processes and
$\tau_1(t)$ the shortest of the characteristic times associated with the
non-dominant processes, with  $\tau_0(t) \ll \tau_1(t)$. Call
$\omega_0=\tau_0^{-1}(\tilde{t})$ and $\omega_1=\tau_1^{-1}(\tilde{t})$,
where $\tilde{t}$ is a given time, arbitrarily chosen. The standard WKB perturbation
method \citep[e.g.,][]{bender,krook} provides the following formal
solution for equation (\ref{eq_rho3}):
\begin{equation}\label{eq_wkb}
 \frac{\rho_1}{\rho_0}(t) = \frac{\rho_1}{\rho_0}(t_0) \exp{ \left\{
\int_{t_0}^{t} \left[ \sum_{n=0}^\infty  
  \omega_0^{1-n}\Ss_n(\tprim) \right] d\tprim\right\}}\;,
\end{equation}
where the $\Ss_n(t)$ are unknown functions to be solved for and $t_0$ is the
initial time. The growth rate of $\rho_1/\rho_0$ is thus given by the term in
brackets within the integrand. If we substitute equation (\ref{eq_wkb}) into
(\ref{eq_rho3}), an equation can be obtained for each order in $\omega_0$.
To calculate it, we split the coefficients $A(t), B(t)$ and $C(t)$ into 
components of order $\omega_0^i$, for $i=0$ through $3$:
\begin{eqnarray}\label{Aterm}
\left\{\matrix{A(t)&=&A_3(t)+A_2(t)+A_1(t)+A_0(t)\;,\nonumber\\
B(t)&=&B_2(t)+B_1(t)+B_0(t)\;,\hfill\\
C(t)&=&C_1(t)+C_0(t)\;.\hfill\nonumber}\right.
\end{eqnarray} 
The general solution for the functions $\Ss_n$ for arbitrary order $n \ge
0$ can be obtained by solving the following algebraic equation:
\begin{eqnarray}\label{eq_omen}
\omega_0^{3-n}\left(\sum_{j=0}^{n}\sum_{m=0}^{n-j} \Ss_j \Ss_m \Ss_{n-j-m} +
3 \sum_{j=0}^{n-1} \dot{\Ss}_j \Ss_{n-1-j} + \ddot{\Ss}_{n-2} \right) + \nonumber \\
 + C_1 \omega_0^{2-n} \left( \sum_{j=0}^{n} \Ss_j \Ss_{n-j} + \dot{\Ss}_{n-1}
\right)
 + C_0 \omega_0^{3-n} \left(\sum_{j=0}^{n-1} \Ss_j \Ss_{n-1-j} 
+ \dot{\Ss}_{n-2} \right)+ \nonumber \\ 
+ B_2 \omega_0^{1-n}\Ss_n + B_1 \omega_0^{2-n}
\Ss_{n-1} + B_0 \omega_0^{3-n} \Ss_{n-2}+ A_{3-n}=0 \;.
\end{eqnarray} 
In this expression, any function with a negative subindex must be considered
equal to zero. Equation ({\ref{eq_omen}}) gives $\Ss_n$ as a function of the
previous orders $\Ss_j$ ($j < n$). The equation for the zero-order solution,
in particular, is as follows:
\begin{equation}\label{eq_ome3_text}
(\omega_0 \Ss_0)^3 + C_1 \,(\omega_0 \Ss_0)^2+ B_2 \;\omega_0 \Ss_0 +
A_3=0\;.
\end{equation}
In the special case that propagation of sound, conduction, and differential
cooling are the dominant processes, equation (\ref{eq_ome3_text}) reduces  to the dispersion relation
of \citet{field65}. 

A few properties of the WKB solution are as follows:

\noindent (1) The asymptotic series (\ref{eq_wkb}) is convergent if
\begin{equation}\label{eq_condit}
|\omega_0^{-n}\Ss_{n+1}(t)|\ll |\omega_0^{1-n}\Ss_n(t)|\;.
\end{equation} 
These conditions are satisfied if $\tau_0(t) \ll \tau_1(t)$ as can be
inferred from the relation $\omega_0^{1-n}\Ss_n(t) \sim {\cal
O}[\omega_0\, (\omega_1/\omega_0)^n]$, where ${\cal O}$ indicates 
the order of magnitude in power of $\omega_1/\omega_0$ . This relation has
been obtained using equation (\ref{eq_omen}) and the equations for the
background evolution. However, the first terms of this series can sometimes 
give an accurate approximation to the exact solution even when $\tau_0(t)$
and $\tau_1(t)$ are of the same order with $\tau_0(t) < \tau_1(t)$.

\noindent (2) The WKB expansion is a global perturbation method, that is, the WKB
solution up to a certain order is good for all times, not just for times much
smaller than the smallest background characteristic time (as would be the
case for a Taylor expansion). 

\noindent (3) We have assumed a uniform background. In spite of this, the criteria we are
going to obtain are applicable locally in a non-uniform background for a
Lagrangian fluid element if
\begin{equation}\label{eq_aplibackuni}
\frac{\lambda}{L_0} \ll \frac{\omega_1}{\omega_0}\ll 1 \;,
\end{equation}
where $\lambda$ is the perturbation wavelength and $L_0$ is the macroscopic
variation length of the unperturbed quantities. That is, the effect of the
non-uniform background is less important than the effect of the non-dominant
processes if condition (\ref{eq_aplibackuni}) holds. An additional condition to apply
the stability criteria in a non-uniform background is: the local uniform
expansion for the Lagrangian fluid element
\begin{equation}
\frac{1}{3}\bf{\nabla \cdot v_0}\it{=\frac{\dot{a}}{a}}\;
\end{equation}
should be much larger than the components of the pure deformation tensor and
the curl of the unperturbed velocity.

In the following sections, we consider various possible cases of dominant and
non-dominant processes (see Table \ref{tabl_cases}).  In section
\ref{sec_intermediate}, intermediate scales, for which sound is one of the
dominant processes, are considered. In section \ref{sother}, extreme scales,
for which sound is a non-dominant process, are considered. This includes
small scales, for which conduction is dominant, as well as long scales, for
which self-gravity, cooling or expansion are dominant.

\begin{table}
\caption{Classification of sections depending on the character of the physical processes
\label{tabl_cases}}
\vspace{0.4cm}
\begin{tabular}{cccccc}
\tableline 
\tableline Section    & Sound & Self--Gravity & Backgr.~expansion & Cooling  & Conduction \\
\tableline \ref{ssou} &  D    &   ND    &  ND          &  ND      & ND         \\
\tableline \ref{s_sougr}&  D  &   D     &  --          &  ND      & ND         \\
\tableline \ref{subsec_shortscale}& ND&         &              &          &  D         \\
\tableline \ref{s_cooling}&ND &         &              &  D       &            \\
\tableline
\end{tabular}
\tablecomments{The physical process can be dominant (D) or non-dominant
(ND). The hyphen in section \ref{s_sougr} indicates that
background expansion must be excluded in that section.}
\end{table}

\section{THE INTERMEDIATE--WAVELENGTH RANGE: SOUND AS A DOMINANT PROCESS} 
\label{sec_intermediate}
At intermediate wavelengths, the propagation of sound can be the dominant
process, either alone or at the same level as one of the other physical
processes.  In the first subsection (sec.~\ref{ssou}), we consider the first
possibility, whereas in the following subsection we study the case in
which sound and self-gravity are simultaneously dominant in the absence of 
background expansion (sec.~\ref{s_sougr}). The readers who are mainly
interested in the thermal stability of the hot interstellar medium, where
self-gravity is negligible, can skip sections  \ref{s_sougr}  and
\ref{sother} and go directly
to section \ref{s_astr} after reading section \ref{ssou}.

\subsection{The Sound Domain}\label{ssou}

When the sound frequency is much larger than the other characteristic
frequencies, there are three WKB solutions: two sound-waves and a
condensation mode, with growth rates for $\rho_1/\rho_0$ which we label
${\Ss_{s}}_{\pm}$ and $\Ss_{c}$, respectively. The solutions to zero
WKB--order, ${\Ss_{0s}}_\pm^{}$ and $\Ss_{0c}$, are
obtained by calculating the three roots of the polynomial 
(\ref{eq_ome3_text}). The coefficients $A_i$, $B_i$ and $C_i$ for this case
are given in Appendix \ref{ApCa}, expressions (\ref{coesd1}). The result is:
\begin{eqnarray}
\omega_0 {\Ss_{0s}}_\pm^{}&=&\pm i\, \omega_s  \;, \label{eq_sds0}\\
\noalign{\vskip -4mm}
\noalign{\hskip 4cm and}
\noalign{\vskip -2mm}
\Ss_{0c} &=& 0 \;. \label{eq_sdc0}
\end{eqnarray}
To zero order, therefore, we have the classical sound waves (\ref{eq_sds0})
and a non-evolving mode (\ref{eq_sdc0}) which corresponds to any perturbation with $v_1=0$, $p_1=0$ and
$\rho_1/\rho_0 \ne 0$. The first order corrections, which we label ${\Ss_{1s}}_\pm$ and
$\Ss_{1c}$, can be calculated using the general WKB solution (\ref{eq_omen})
for $n=1$ (and the coefficients $A_i$, $B_i$ and $C_i$ of
eq.~[\ref{coesd2}]). The results are given in the following separately
for the condensation mode and for the sound waves.

\subsubsection{The Condensation Mode}\label{sec_condens_mode}
The first order correction for the condensation mode is
\begin{equation}\label{eq_sdc1}
\Ss_{1c}=-\frac{(\gamma-1)}{\gamma}\frac{\mu}{\kboltz}\left\{ T_0
\left[\frac{\partial}{\partial T}
\left(\frac{\Ll}{T}\right) \right]_p +\frac{\chi_0}{\rho_0}\frac{k^2}{a^2}\right\}\;.
\end{equation} 
Together with $\Ss_{0c} =0$, the foregoing indicates that this mode does not propagate. 
Conduction has a stabilizing effect, whereas the effect of cooling depends on
the sign of $[\partial (\Ll/T)/\partial T]_p$. The reason for the appearance
in equation (\ref{eq_sdc1}) of the partial derivative of $\Ll/T$ {\it at
constant pressure} is that pressure remains almost spatially uniform in the
condensation mode. Indeed, the solution for pressure and velocity is:
\begin{equation}
\begin{array}{rclcl}
\fracdps{v_1}{c_s}&=&i \fracdps{\Ss_{1c}}{\omega_s}\fracdps{\rho_1}{\rho_0}
 &\sim & {\cal O}\left(\fracdps{\omega_1}{\omega_0}\right)\;
 \fracdps{\rho_1}{\rho_0}\;,\\
\noalign{\vspace{5mm}}
\fracdps{p_1}{p_0}&=&-\gamma\fracdps{\dot{\Ss}_{1c}+ 2
(\dot{a}/a)\Ss_{1c}+\Ss_{1c}^2-\omega_g^2}{\omega_s^2}
\fracdps{\rho_1}{\rho_0} &\sim & {\cal O}\left(\fracdps{\omega_1^2}{\omega_0^2}\right)\;
\fracdps{\rho_1}{\rho_0}  \;,
\end{array}
\label{eq_relsizesound}
\end{equation}
where ${\cal O}$ indicates the order of magnitude in power of
$\omega_1/\omega_0$, with $\omega_1$ in this case given by the shortest of the conduction and cooling times
appearing in equation (\ref{eq_sdc1}). Hence, the relative pressure perturbation
obtained from the linear system is zero down to second WKB order.

The second order WKB correction for the growth rate of $\rho_1/\rho_0$ for 
the condensation mode is equal to zero as can be shown using equation
(\ref{eq_omen}) for $n=2$: the following non-zero correction appears
at third WKB order.

\subsubsection{The Sound Modes}\label{sec_sound_mode}
The first order WKB correction for the sound modes is:

\vspace{-5mm}
\begin{equation}\label{eq_sds1}
{\Ss_{1s}}_\pm^{}=-\frac{(\gamma-1)}{2}\frac{\mu}{\kboltz} \left\{
\frac{\gamma-1}{\gamma} \left[ \left( \frac{\partial \Ll}{\partial T}
\right)_S + \frac{\chi_0}{\rho_0}\frac{k^2}{a^2} \right]
-\left( \frac{3}{2}-\frac{1}{\gamma} \right) \frac{\Ll_0}{T_0}  \right\} 
+\frac{(3\gamma -5)}{4}\frac{\dot{a}}{a} \quad,
\end{equation}

\noindent where $S$ stands for the specific entropy. Since $\Ss_{0s}$ has no real part, the
change in amplitude is given by $\Ss_{1s}$.  From equation (\ref{eq_sds1}) we see
that a net background radiative cooling ($\Ll_0 > 0$) has a destabilizing
effect. Expansion tends to destabilize these modes if $\gamma > 5/3$. Conduction
has an unconditionally stabilizing effect for the sound mode.
In contrast to the condensation mode (see eq.~[\ref{eq_sdc1}]), now the
$T$-derivative of $\Ll$ that appears in the growth rate is {\it at constant entropy}. This is easily
understandable, since entropy is changed by conduction and cooling only,
which are non-dominant processes in the present case. So, these modes evolve
essentially at constant entropy. Self-gravity does not appear in this first
order correction (assuming that all the non-dominant processes have a 
characteristic time of the same order).
Putting together the zeroth- and first order expression, we can write
the WKB solution for the density perturbation in the sound mode as:
\begin{equation}\label{eq_sdss}
\frac{\rho_1}{\rho_0}  \propto \; \fracdps{1}{T_0^{1/4} a^{1/2}}\;\exp\left\{ \pm i \int_0^t \omega_s
dt - \frac{(\gamma-1)^2}{2 \gamma}\frac{\mu}{\kboltz} \int_0^t
\left[ T_0 \left( \frac{\partial}{\partial T}
\frac{\Ll}{T}  \right)_S + \frac{\chi_0}{\rho_0}\frac{k^2}{a^2}  \right] dt \right\}\;.
\end{equation}

The solution for the pressure and density up to first WKB order for the sound
waves is:
\begin{eqnarray}\label{eq_Apsou_soundv_1}
\frac{v_1}{c_s}&=&\frac{i(\omega_0
\Ss_{0s}+\Ss_{1s})}{\omega_s}\frac{\rho_1}{\rho_0}\;,\\
\label{eq_Apsou_soundp_1}
\frac{p_1}{p_0}&=&\gamma\left[1\mp
\frac{i}{\omega_s}\left(\frac{\Omcdp + \tchi^{-1}}{\gamma}  - \Omcdrho -
\tchi^{-1}  \right)\right] 
\frac{\rho_1}{\rho_0}\;.
\end{eqnarray}
To zero-order, this reduces to the classical $\rho_1/\rho_0 = (1/\gamma)
(p_1/p_0) = \mp v_1/c_s $ of the elementary sound waves.  

\subsubsection{Link with the Previous Literature: the Generalized Field Length}
\label{sec_appl_conds}

The solutions of the previous subsections generalize those obtained by
previous authors for particular cases. Thus, in the absence of background
expansion and with no net background cooling (i.e., when there is 
thermal equilibrium in the unperturbed state), our WKB solutions
(\ref{eq_sds0}), (\ref{eq_sdc1}) and (\ref{eq_sds1}) reduce to the solutions
of \citet{field65}. In the absence of thermal conduction, for $\gamma=5/3$
and assuming that $\Ll/\rho$ only depends on $T$, our solution for the
condensation mode reduces to that of \citet{mathews}. In the absence of
thermal conduction and for $\gamma=5/3$, our solutions reduce to those of \citet{balbus86}.

For the temperatures and densities for which cooling is destabilizing, \citet{field65} 
found that for each mode, sound and condensation, there
exists a critical length, which we will call henceforth {\it classical
Field length}, defined as the maximum wavelength for which conduction can
suppress the instability. Field considered a static background in thermal
equilibrium. For a background undergoing cooling and expansion, 
we can then now define a {\it generalized Field length} (and the
corresponding {\it generalized Field wavenumber}, $\kcrit$) for each mode as the
wavelength (or wavenumber) separating stability domains.
The generalized Field wavenumber for the condensation mode is thus given by:
\begin{equation}\label{eq_kcrc}
\left(\frac{\kcrit}{a}\right)^2=- T_0 \frac{\rho_0}{\chi_0}\left[\frac{\partial}{\partial T}
\left(\frac{\Ll}{T}\right) \right]_p\;.
\end{equation}
This expression has been obtained setting equation (\ref{eq_sdc1}) equal to
zero. Note that cooling must have a destabilizing effect ($[\partial(\Ll/T)/\partial
T]_p<0$) for the existence of this critical length.
There is also a generalized Field length for the sound waves, given as the
positive real solution of equation $\Ss_{1s}=0$ which usually is 
larger than the one for the condensation mode.

\subsubsection{Comparison of WKB and Exact Solutions for a Particular Case}

In this section, we consider an idealized case to compare the WKB solutions
and the numerical solutions of the system (\ref{eq_rho1})-(\ref{eq_p1}) (and
leave for section \ref{s_astr} the application to a realistic ISM or IGM). We consider a
completely ionized plasma of pure hydrogen with temperature above $10^5$K. 
For those temperatures, cooling is almost only due to bremsstrahlung because the line and
recombination emission due to elements heavier than hydrogen is not
present. Conduction, in turn, is due almost exclusively to electrons. Thus,
$\Ll$ and $\chi$ can be taken in the form \citep{rybicki,spitzer}:
\begin{eqnarray}
\Ll &=& 10^{-3} n_H T^{1/2}\;\hbox{erg gr$^{-1}$s$^{-1}$}\;,\label{eq_bremss}\\
\noalign{\vskip 3mm}
\chi &=& \frac{1.84 \times 10^{-5}}{\log{\LambdaCoulomb}} T^{5/2}\;\hbox{erg s$^{-1}$K$^{-1}$cm$^{-1}$}\;,\;
\hbox{with}\;\log{\LambdaCoulomb}=29.7+\log{\left(T\times 10^{-6}/{n_e}^{1/2}\right)}\;,
\label{eq_condspit}
\end{eqnarray}
where the averaged Gaunt factor has been chosen equal to $1.2$, $n_H$ is the
proton number density in cm$^{-3}$, $n_e$ is the electron number density in
cm$^{-3}$, and $T$ is measured in K. In this subsection, the Coulomb
logarithm $\log \LambdaCoulomb$ is set equal to $29.7\,$.  The cooling
function (\ref{eq_bremss}) has a destabilizing effect on the condensation
mode (see eq.~[\ref{eq_sdc1}]) and on the sound waves (see
eq.~[\ref{eq_sds1}]).  We assume a non-expanding background, so that $\rho_0$
remains constant in time. The evolution of the background temperature can be
easily obtained from equations (\ref{eq_ener0}) and (\ref{eq_bremss}).

Using as time unit the period of a sound wave at time $t=0$, ${\tau_s}_0^{}$,
and equations (\ref{eq_bremss}) and (\ref{eq_condspit}) for cooling and
conduction, it turns out that the coefficients of the relative perturbations 
in the system (\ref{eq_rho1})-(\ref{eq_p1}) depend on $k$
and $\rho_0$ through the single parameter $\rho_0/k$ only (equivalently, 
$\lambda\; n_H$). This allows a large simplification in the presentation of 
the results in the following.

\begin{figure} 
\plotone{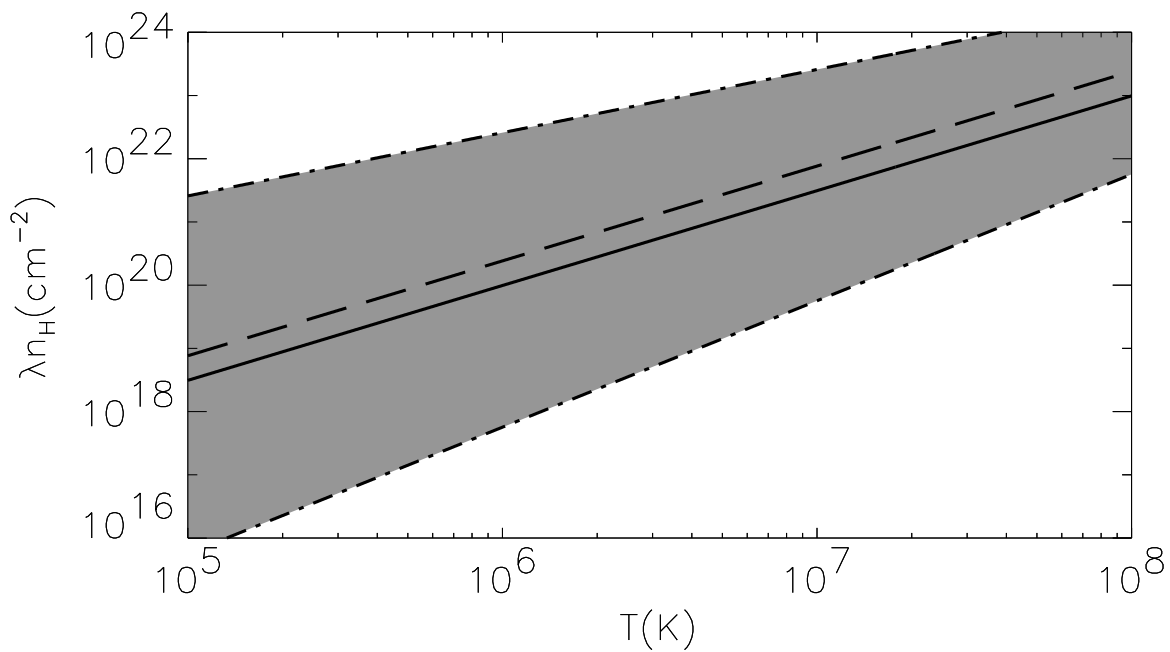}
\caption{Critical wavelengths and boundaries of the sound domain.
The solid line indicates the critical wavelength for the
condensation mode. The dashed line corresponds to the critical wavelength for
the sound waves. The two dash--dotted lines are the points where 
$\tau_c^{-1}=\omega_s$ (upper straight) and $\tau_\chi^{-1}=\omega_s$ 
(lower straight): the {\it sound domain}, therefore, is contained
within the shaded band. \label{fig_3_1a}}
\end{figure}

\noindent {\bf (a) Critical wavelengths.} The critical wavelength for the
condensation mode (i.e., the generalized Field length) is obtained from
equation (\ref{eq_kcrc}), whereas for the sound waves it can be calculated by
setting the growth rate (\ref{eq_sds1}) equal to zero.  Both are represented
in Figure \ref{fig_3_1a} as implicit functions of $\lambda n_H$ and
$T_0$. The critical wavelength for the condensation mode (solid line), is
smaller than the critical wavelength for the sound waves (dashed
line). Figure \ref{fig_3_1a} also shows the region (drawn as a shaded band)
where $\tchi^{-1} < \omega_s$ and $\omega_s > \tau_c^{-1}$. The {\it sound
domain} is contained within that band, but, as seen just below, the WKB
solutions of the previous subsections are good approximations also close to
the boundaries of the region. Well above the shaded band, cooling acts on the
fastest timescale; well below it, conduction is dominant.

\begin{figure} 
\plotone{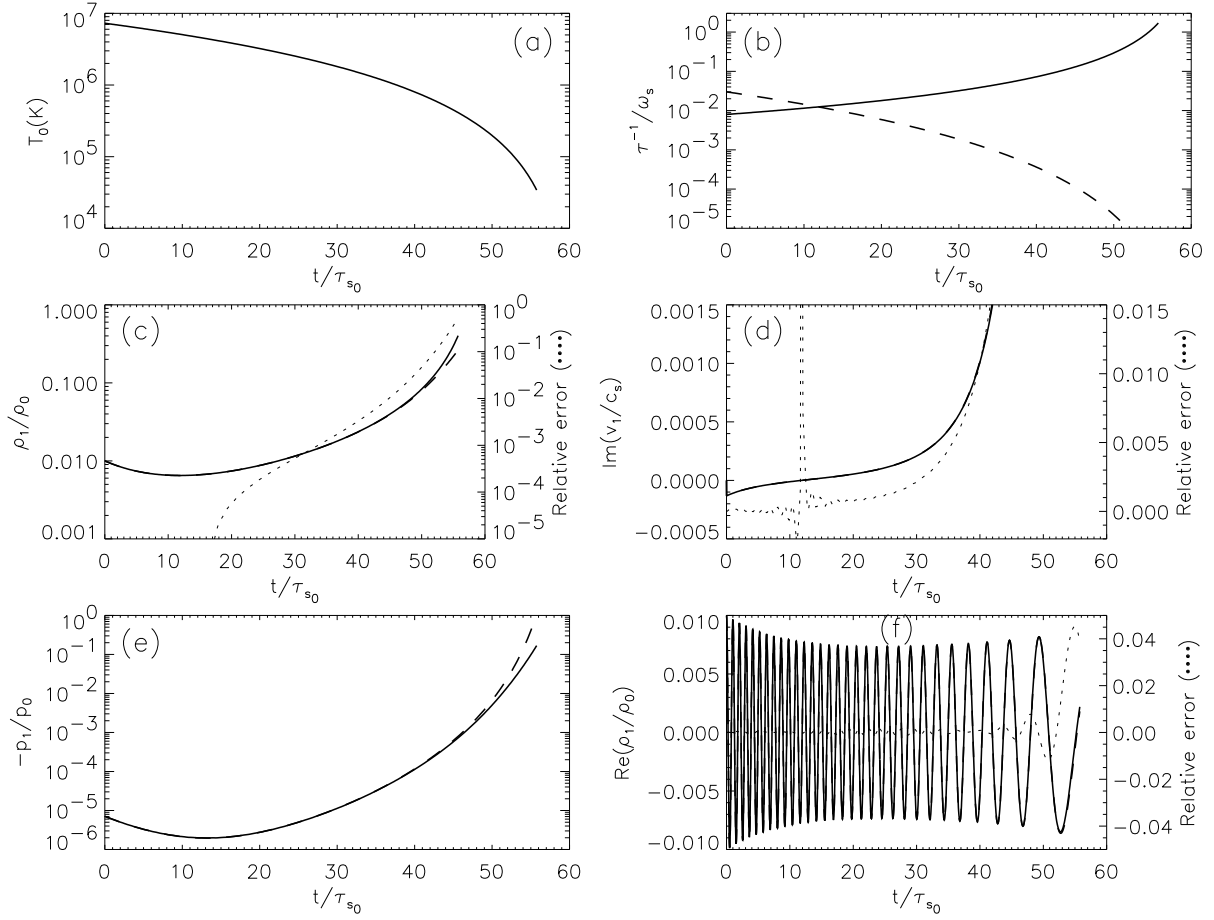}
\caption{Comparison of the WKB and the numerical solutions for the
condensation and sound modes. (a) Evolution of the background
temperature. (b) Solid line: time evolution of $|\Omcdp|/\omega_s$.  Dashed
line: evolution of $\tau_\chi^{-1}/\omega_s$. (c), (d), and (e) show the
evolution of the condensation mode: the solid line is the WKB solution and
the dashed line is the numerical solution (in panel d, they coincide within
the resolution of the figure).  Panel (f) shows the evolution of the density
perturbation for a sound wave (again, WKB and numerical solution coincide
in the figure). Panels (c), (d) and (f) also show, as a dotted line, the
relative error between the numerical and the WKB solutions. \label{fig_3_1b}}
\end{figure}

\noindent {\bf (b) Comparison of WKB- and exact solutions.} We consider both
the condensation mode and the sound waves for a case defined by $\lambda
{n_H}=10^{21}\,$cm$^{-2}$ and $T_0=7.3 \times 10^6\,$K at $t=0$
(Fig.~\ref{fig_3_1b}). With this choice, both modes have initially a
wavelength below the critical value. Figures \ref{fig_3_1b}a and b show the
time evolution of the background temperature and of $|\Omcdp|/\omega_s$ and
$\tau_\chi^{-1}/\omega_s$. The other four panels in the figure compare the
WKB solution with the numerical solution of the linear system
(\ref{eq_rho1})-(\ref{eq_p1}) for an initial amplitude of the perturbation
$\rho_1/\rho_0=0.01$.  Panels (c), (d) and (e) show the time evolution of the
condensation mode (solid: WKB approximation; dashed: numerical solution)
and the relative error between them (dotted line). The WKB solution turns out to be an
excellent approximation to the exact solution: both curves are almost
indistinguishable except when $|\Omcdp| \sim \omega_s$. In fact, the relative
error of the WKB solution for $\rho_1/\rho_0$ is still below $0.01$ when
$|\Omcdp|/\omega_s$ reaches the value $0.1$.  Initially, the condensation
mode has a wavelength below the critical value, but the latter decreases
as the system cools (see Fig.~\ref{fig_3_1a}) and their ratio grows 
above unity for $t/{\tau_s}_0^{} = 12$.  Therefore, $\rho_1/\rho_0$ first
decreases, then reaches a minimum at that time, and grows thereafter.
Figure \ref{fig_3_1b}f shows the WKB and numerical solutions for the sound
wave. Again, the agreement between both solutions is excellent.  The
wavelength ratio follows a similar evolution as for the condensation mode,
and so does the perturbation amplitude.

\subsection{Solutions on Scales around the Jeans Length: the
Acoustic--Gravity Domain without Background Expansion}
\label{s_sougr}

Going to spatial scales larger than in the foregoing section, there is
interesting new physics when the wavelength approaches the Jeans length,
$L_J=2 \pi c_s/\omega_g\,$, that is, as soon as $\omega_g$ is no longer small
compared to $\omega_s$.  In this subsection we study the regime in which
either sound or gravity (or both simultaneously) are dominant and the
background is slowly cooling or heating but with zero expansion rate 
(if $\dot{a} \ne 0$, the expansion would become as dominant as
self-gravity itself after a time of order $\tau_g$ at the latest, see
eq.~[\ref{eq_mom0}], and the WKB model could not be used).

Consider, then, the case in which $|\Omgc|$, $\tau_c^{-1}$ and $\tchi^{-1}$
are all much smaller than both $\omega_s$ and $\omega_g$.  There are then
three WKB solutions: two acoustic--gravity modes and a condensation mode. We
use the symbols ${\Ss_{ag}}_\pm$ and $\Ss_c$ for their respective growth
rates for $\rho_1/\rho_0$.  As in the previous case, the zero-order solutions
are obtained by calculating the three roots of the polynomial
(\ref{eq_ome3_text}), but now the coefficients $A_i$, $B_i$ and $C_i$ have
the values given in Appendix \ref{ApE}, expressions (\ref{coesgd1}). The
result is: 
\begin{eqnarray}
\omega_0 {\Ss_{0ag}}_\pm^{}&=&\pm i \sqrt{\omega_s^2-\omega_g^2} \;\label{eq_sgds0}\\
\noalign{\vskip -4mm}
\noalign{\hskip 3.6cm and}
\noalign{\vskip -2mm}
\Ss_{0c} &=& 0 \;. \label{eq_sgdc0}
\end{eqnarray}
To zero order, therefore, we have a couple of acoustic--gravity modes
(eq.~[\ref{eq_sgds0}]) like those obtained by \citet{jeans} and a non-evolving mode
(eq.~[\ref{eq_sgdc0}]). The latter corresponds to any perturbation with $v_1=0$ and
$p_1/p_0=\gamma (\omega_g^2 / \omega_s^2) \rho_1/\rho_0$, i.e., equilibrium
between pressure gradient and self-gravity. If $\omega_g<\omega_s$ the
acoustic--gravity modes are waves. They have a frequency smaller than a pure
sound wave with the same wavelength because the self-gravity force acts
against the pressure gradient force. If $\omega_g>\omega_s$ these modes grow
or decay exponentially. In the growing mode, in particular, the perturbation 
collapses gravitationally and the pressure gradient is not large enough to
prevent it. The introduction of cooling and
thermal conduction as non-dominant processes makes the amplitude of these
modes change. These modifications are given by the WKB corrections of order
one and greater.

To calculate the first order corrections for the growth rates of
$\rho_1/\rho_0$ for those modes, namely ${\Ss_{1ag}}_\pm$ and $\Ss_{1c}$, we turn again to the 
general WKB solution (eq.~[\ref{eq_omen}]), using $n=1$ and the coefficients given
in expressions (\ref{coesgd2}) of Appendix \ref{ApE}. We show and discuss them in the following:

\subsubsection{Condensation Mode}\label{subsec_condgrav}

The first order WKB correction for the condensation mode is given by
\begin{equation}\label{eq_sgdc1}
\Ss_{1c}=-\frac{(\gamma-1)}{\gamma}\frac{\mu}{\kboltz}\left\{ T_0
\left[\frac{\partial}{\partial T}
\left(\frac{\Ll}{T}\right) \right]_p - \gamma \frac{\omega_g^2}{\omega_s^2}
\left(\frac{\partial \Ll}{\partial T} \right)_{\rho}
+\frac{\chi_0}{\rho_0}\frac{k^2}{a^2} \left( 1 - \gamma
\frac{\omega_g^2}{\omega_s^2} \right)\right\}\left( 1 - \frac{\omega_g^2}{\omega_s^2}
\right)^{-1}\;.
\end{equation}
This mode reduces to the sound domain condensation mode (\ref{eq_sdc1}) when
$\omega_g \ll \omega_s$. We introduce the quantity
\begin{equation}\label{eq_defF}
F = \frac{p_1}{p_0}-\gamma\frac{\omega_g^2}{\omega_s^2} \frac{\rho_1}{\rho_0} \;,
\end{equation} 
which is proportional to the perturbation of the net volume force (pressure
gradient minus self-gravity). Using this quantity, the relative perturbations
can be written
 \begin{equation}\label{mquan}
 \frac{v_1}{c_s}=\frac{i\Ss_{1c}}{\omega_s} \; \frac{\rho_1}{\rho_0}\sim {\cal
 O} \left( \frac{\omega_1}{\omega_s} \right) \;
 \frac{\rho_1}{\rho_0}\;,\qquad F = -\gamma
\frac{\dot{\Ss}_{1c}+\Ss_{1c}^2}{\omega_s^2}  \frac{\rho_1}{\rho_0} \sim 
 {\cal O} \left(\fracdps{\omega_1^2}{\omega_s^2}\right) 
\frac{\rho_1}{\rho_0}\;,
 \end{equation} 
where $\omega_1$ is the inverse time-scale of the non-dominant processes
(conduction and cooling). This is similar to equation (\ref{eq_relsizesound}) 
but  now  the sound-gravity  modes,  that propagate  or  grow much  faster  
than the  condensation  mode,  lead to  the cancellation of  the gravity
force  by the pressure  gradient instead of  to a condition of uniform pressure throughout.

The first order correction (\ref{eq_sgdc1}) shows three interesting
features. First, it diverges when $\omega_s=\omega_g$, i.e., 
there is a turning-point in the WKB sense. The turning point can easily be
reached in a cooling and non-expanding medium for wavelengths for which,
initially, $\omega_s > \omega_g$: the former frequency decreases with time
whereas the latter stays constant. As known from the WKB theory, when a
solution goes through a turning point, a mixing of modes takes place.
Second, conduction has a destabilizing effect in the range 
$\gamma \omega_g^2 > \omega_s^2 > \omega_g^2\;$. This can be explained using
the equation of evolution for the perturbed entropy, $S_1$, and the relation
between $S_1$ and $T_1/T_0$ following the condition of permanent 
equilibrium between gravity and pressure gradient.
Third, the combination of partial derivatives
of $\Ll$ is quite different to the corresponding expression in equation
(\ref{eq_sdc1}). In the following we explain the origin of the last feature.

To explain the origin of the cooling terms in equation (\ref{eq_sgdc1}) we
can use the results of \citet{balbus86}. Using Lagrangian
perturbations in the entropy equation, he derived an instability
criterion for the condensation mode in the absence of conduction in the
particular case in which the perturbation of an arbitrary thermodynamic
variable $A$ is kept equal to zero during the evolution of the condensation
mode. The resulting criterion is obtained from the following equation:
\begin{equation}\label{entrlag}
\frac{1}{\Delta S}\frac{d \Delta S}{d t}=-\left[\frac{\partial }
{\partial S}\left(\frac{\Ll}{T}\right)\right]_A\;,
\end{equation}  
where $\Delta$ indicates Lagrangian perturbation. \citet{balbus86} applies
this equation to the cases $A=p$ and $A=\rho$. In the present section,
in turn, the physical quantity whose perturbation remains equal to zero is $F$
(eq.~[\ref{mquan}]); on the other hand, Lagrangian and
Eulerian perturbations are indistinguishable since the background is uniform
\citep[e.g.,][]{shapiro}. It can then be shown that the cooling terms in
equation (\ref{eq_sgdc1}) can be derived from the right-hand side of
equation (\ref{entrlag}), using the relationship between the growth rates of $S_1$
and $\rho_1/\rho_0$.

Critical lengths are computed solving the equation $\Ss_{1c}=0$. This
equation can be expressed as:
\begin{equation}\label{eq_critcondag}
\frac{\lambcrit^4}{L_J^4} + \left( \frac{\tauchiJ^{-1}}{\OmcdrhoF}-
\frac{1}{\gamma}\frac{\Omcdp}{\OmcdrhoF} \right)
\frac{\lambcrit^2}{L_J^2} - 
\frac{1}{\gamma} \frac{\tauchiJ^{-1}}{\OmcdrhoF} = 0
\end{equation}
where $\lambcrit$ is the critical wavelength, $\tauchiJ^{-1}$ is
$\tau_\chi^{-1}$ evaluated for $\lambda=L_J$, and 
\begin{equation}\label{eq_omcrhof}
\OmcdrhoF = (\gamma-1)\frac{\mu}{\kboltz} \left(\frac{\partial \Ll}{\partial
T}\right)_\rho \;,
\end{equation}
i.e., minus the growth rate due to isochoric cooling in the classical theory of \citet{field65}.
At the Jeans length, it is natural to expect cooling to be much faster than
conduction ($\tauchiJ^{-1} \ll \OmcdrhoF$). Assuming this, the
solutions of equation (\ref{eq_critcondag}) are given by: 
\begin{equation}
{\lambcrit^2}_1= - \frac{\tauchiJ^{-1}}{\Omcdp} L_J^2 \qquad \hbox{and}
\qquad {\lambcrit^2}_2 = \frac{1}{\gamma}\frac{\Omcdp}{\OmcdrhoF} L_J^2 \;.
\end{equation}
For the existence of a critical length, the right-hand side of the corresponding
expression must be positive. The first critical length coincides with that of
equation (\ref{eq_kcrc}), which corresponds to a condensation mode in the
sound dominated case. The second critical length is new, and has a value that
is near $L_J$. If both critical lengths exist and
$\lambcrit_{1,2}/L_J<1$, the condensation mode will be unstable
for wavelengths $\lambda$ such that $\lambcrit_1 < \lambda < \lambcrit_2$.

\subsubsection{Acoustic--Gravity Modes}\label{subsec_acgrav}

The first order WKB correction for the acoustic--gravity modes is
\begin{equation}\label{eq_sgds1}
{\Ss_{1ag}}_\pm=-\frac{(\gamma-1)}{2}\frac{\mu}{\kboltz}\left\{
\frac{\gamma-1}{\gamma} \left[\left( \frac{\partial \Ll}{\partial T}\right)_S + 
\frac{\chi_0}{\rho_0}\frac{k^2}{a^2}  \right]
-\left(\frac{3}{2}-\frac{1}{\gamma}\right) \frac{\Ll_0}{T_0}  \right\}
\left( 1 -  \frac{\omega_g^2}{\omega_s^2} \right)^{-1}\;.
\end{equation}
This term coincides with the first-order correction for the sound waves given
in equation (\ref{eq_sds1}) for zero background expansion but divided by $1 -
\omega_g^2/\omega_s^2$.  For $\omega_g < \omega_s$, this is not surprising
because the acoustic--gravity waves behave in that case like sound waves of
the same wavelength but with a smaller frequency.  
Similarly as for the sound domain, for this mode
there is only one critical wavelength, obtained solving $\Ss_{1ag}=0$.

\section{THE SHORT-- AND LONG--WAVELENGTH RANGES}
\label{sother}

For completeness (and later use in section \ref{s_astr}), we briefly discuss
in this section two results for perturbations in the conduction and cooling
domains.

\subsection{Small Spatial Scale: Heat Conduction Domain}
\label{subsec_shortscale}

At small enough spatial scale  (a) thermal conduction is the dominant process
and (b) $\omega_s^{-1}$ is much smaller than $\tau_g$, $\tau_e$ and $\tau_c$,
so that $\tchi \ll  \omega_{\rm{s}}^{-1} \ll \min(\tau_g, \;\tau_e,
\;\tau_c)$.  Two simple consequences of the foregoing are: (1) conduction can
eliminate any temperature  gradients well before the sound  waves can cross a
wavelength  and (2) the evolution of the dominant processes (conduction and
sound propagation) occurs without important evolution of the background
through cooling and expansion. The solutions to zero and first
WKB order  must thus  coincide with  the solutions  obtained  by \citet{field65},
viz.~a condensation  mode at zero order and  quasi-isothermal sound waves at
first order. The condensation mode evolves on the timescale of the 
thermal conduction, $\tchi$. Using this fact in equations
(\ref{eq_rho1})--(\ref{eq_p1}) one obtains $v_1/c_s \sim {\cal O} (
\omega_s/\tau_\chi^{-1} ) \;p_1/p_0 \;$, $\rho_1/\rho_0 \sim {\cal O}
(\omega_s^2/\tau_\chi^{-2}) \;p_1/p_0$ and
\begin{equation}\label{eq_p_timescale}
\fracdps{d}{dt}\left(\frac{p_1}{p_0}\right) = - \fracdps{1}{\tchi} \fracdps{p_1}{p_0}
    \quad.
\end{equation}
The condensation mode at small length-scales is therefore completely
stabilized by conduction. As a consequence of the smallness of
$\rho_1/\rho_0$, to obtain the WKB solutions for this mode it is necessary to
use a third order differential equation for $p_1/p_0$ instead of the
differential equation (\ref{eq_rho3}) for $\rho_1/\rho_0$.

\subsection{Large Spatial Scale: Cooling Domain}
\label{s_cooling}

At large enough wavelength,  cooling, expansion and/or self-gravity dominate,
so  that  the  following   ordering  of  timescales  applies:  $\min{(\tau_g,
\;\tau_e,\;\tau_c)} \ll \omega_{\rm{s}}^{-1} \ll \tchi $. Whenever cooling or
expansion dominate,  no WKB solutions can  be calculated, since  in that case
background  and perturbation  evolve on  the same  or  comparable timescales.
However,  a zero  order  solution can  be  obtained by  direct inspection  of
equations  (\ref{eq_rho1})  through  (\ref{eq_p1})  when there  is  a  single
dominant process. For instance, if cooling is the single dominant process, we
expect   the  zero  order   solution  to   have  a   growth  rate   of  order
$\tau_{mc}^{-1}=\max{(\tau_c^{-1},\Omgc)}$  for  all  relative  perturbations
$\rho_1/\rho_0$,  $v_1/c_s$  and  $p_1/p_0$.  Inserting that  timescale  into
equations  (\ref{eq_rho1})-(\ref{eq_p1}) one  finds: $v_1/c_s  \sim  {\cal O}
(\omega_s/\tau_{mc}^{-1})\;p_1/p_0$   and   $\rho_1/\rho_0   \sim  {\cal   O}
(\omega_s^2/\tau_{mc}^{-2})\;p_1/p_0$.  This perturbation  evolves  at almost
constant density  because the  propagation of sound  is much slower  than the
growth by  cooling. The growth  rate at zero  order can be obtained  from the
previous considerations and equations (\ref{eq_p1}) and (\ref{eq_coe4}):
\begin{equation}\label{eq_cooldo}
\frac{d}{dt} \left(\frac{p_1}{p_0}\right) = -\Omcdrho \; \left(\fracdps{p_1}{p_0}\right)\;.
\end{equation}
\noindent \citet{balbus86} obtained this growth rate for the perturbed entropy using 
equation (\ref{entrlag}) with $A=\rho$.

\section{GENERALIZED FIELD LENGTH IN AN ASTROPHYSICAL COOLING MEDIUM}
\label{s_astr}

\begin{figure} 
\plotone{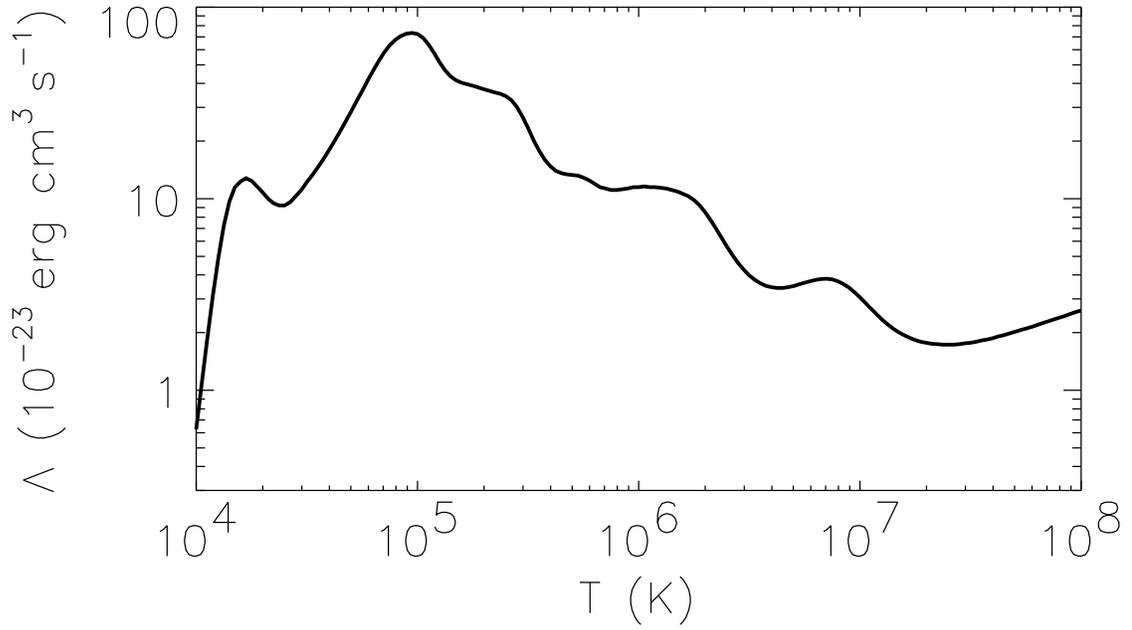}
\caption{Cooling function $\funcool (T)$ of J. C. Raymond (2001, private
communication) for standard abundances. \label{fig_apl1}}
\end{figure}

We consider the thermal stability of an optically thin astrophysical medium
at a temperature in the range $10^4-10^8$K which is undergoing net
cooling. The medium is assumed to be uniform (or only weakly non-uniform in
the sense of eq.~[\ref{eq_aplibackuni}]). Old SNRs and super-bubbles are
examples of such a medium. The generalized Field length, which can be
computed using the results of section \ref{sec_appl_conds}, separates the
stable regions from the unstable ones in wavenumber Fourier space. 
Self-gravity can be ignored because for the considered medium the Jeans
length is usually much larger than the Field one. Assuming no heating and 
permanent ionization equilibrium, the cooling function only depends on the 
instantaneous properties of the medium and can be expressed as follows:
\begin{equation}\label{eq_coolthin}
\Ll= \left( \frac{n_e n_H}{\rho^2} \right) \rho \, \funcool(T)\;,
\end{equation}
where $n_e$ is the electron number density, $n_H$ is the total number density
of hydrogen (atoms and ions) and $\funcool$ is a function that only depends
on temperature. We use the function $\funcool (T)$ provided by J. C. Raymond
(2001, private communication) for standard abundances and $n_H=1\,$cm$^{-3}$ 
(see Fig.~\ref{fig_apl1}), which is an
improved version of the cooling function of \citet{raymond76}. For $T<3
\times 10^4\,$K, the function $\funcool (T)$ has a very weak dependence on
$n_H$, which can be safely neglected in the present context. In the considered
temperature range, $\gamma = 5/3$ and the thermal conduction coefficient, $\chi$, is given by
equation (\ref{eq_condspit}). A number of authors have
suggested a substantial reduction of $\chi$ in the
Intergalactic Medium due to, for instance, kinetic instabilities driven by
temperature gradients \citep[e.g.,][]{pistinner96}, a tangled magnetic field,
which makes the effective mean free path of electrons smaller
\citep[e.g.,][]{rosner89} and may cause magnetic mirroring
\citep[e.g.,][]{chandran99}. Thus, as is usual, we include a constant factor $\xi$,
with value between $0$ and $1$, multiplying the conduction coefficient $\chi$.

\subsection{Unmagnetized Background}\label{ss_unmag_backg}

The generalized Field length, $\lambcrit$, can be calculated inserting
the cooling function (\ref{eq_coolthin}) into the growth rates
(\ref{eq_sdc1}) and (\ref{eq_sds1}) for the condensation and sound modes,
respectively, and setting them equal to zero. The result is:
\begin{equation}\label{eq_kcriapli}
\lambcrit^2 \, n_H^2 \, x \,=\,  4 \pi^2 \, \xi \,\chi \,
\frac{T}{\funcool (T)}\, \left[\alpha - \frac{d \log{\funcool}}{d \log{T}} \right]^{-1}\;, 
\end{equation}
with $\alpha = 2$ for the condensation mode and $\alpha= 3/4$ for the sound
waves. In equation (\ref{eq_kcriapli}), $x=n_e/n_H$ and all quantities are evaluated in the background.
If the logarithmic temperature gradient of $\funcool$ is
positive and steep enough, the right-hand side of equation (\ref{eq_kcriapli}) becomes
negative and there is no critical length: the corresponding mode is stable
at that temperature for all wavelengths.

\begin{figure} 
\plotone{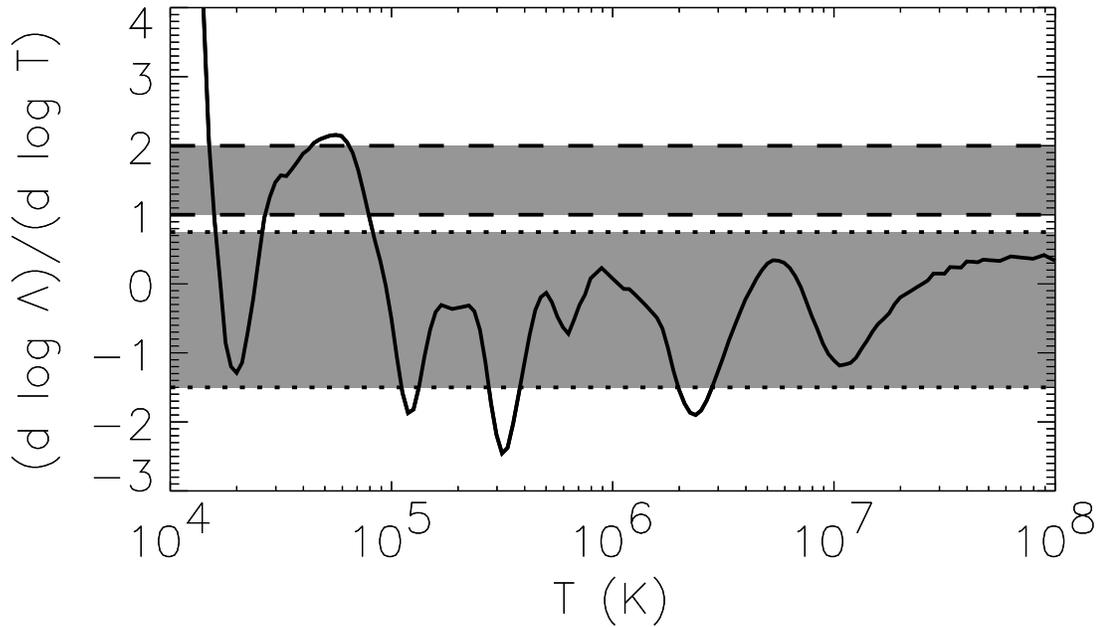}
\caption{The solid line shows $(d \log{\funcool})/(d \log{T})$ for the
function $\funcool(T)$ represented in Figure \ref{fig_apl1}. The horizontal
lines give the upper limits of existence of a critical length for the
condensation mode (dashed) and for the sound waves (dotted), whereby in both
cases the lower line corresponds to the Field problem and the upper one to
the case with net background cooling.  Wherever the solid line falls onto the
shaded bands, the system with net background cooling is unstable whereas the
results for thermal equilibrium predict stability.
\label{fig_apli2}} 
\end{figure}

It is instructive to compare equation (\ref{eq_kcriapli}) with the classical
Field length (sec.~\ref{sec_appl_conds}), i.e., the critical length if the
background were in thermal equilibrium through the action of a heating which
exactly compensates the cooling.  Since we are interested primarily in the
destabilizing effect of the  cooling, we assume a density-- and temperature--independent
heating.  The classical Field length is also given by equation
(\ref{eq_kcriapli}) but with $\alpha = 1$ for the condensation mode and
$\alpha = -3/2$ for the sound waves. From these values of $\alpha$, comparing
them with the values obtained above (see Fig.~\ref{fig_apli2}), we gather
that there are more (and wider) unstable temperature ranges in the case with
net background cooling than in the classical Field problem.  In fact, the
difference is quite striking for the sound waves, which change from stable to
unstable almost everywhere if one suppresses the heating providing thermal
equilibrium.

\begin{figure} 
\epsscale{0.8}
\plotone{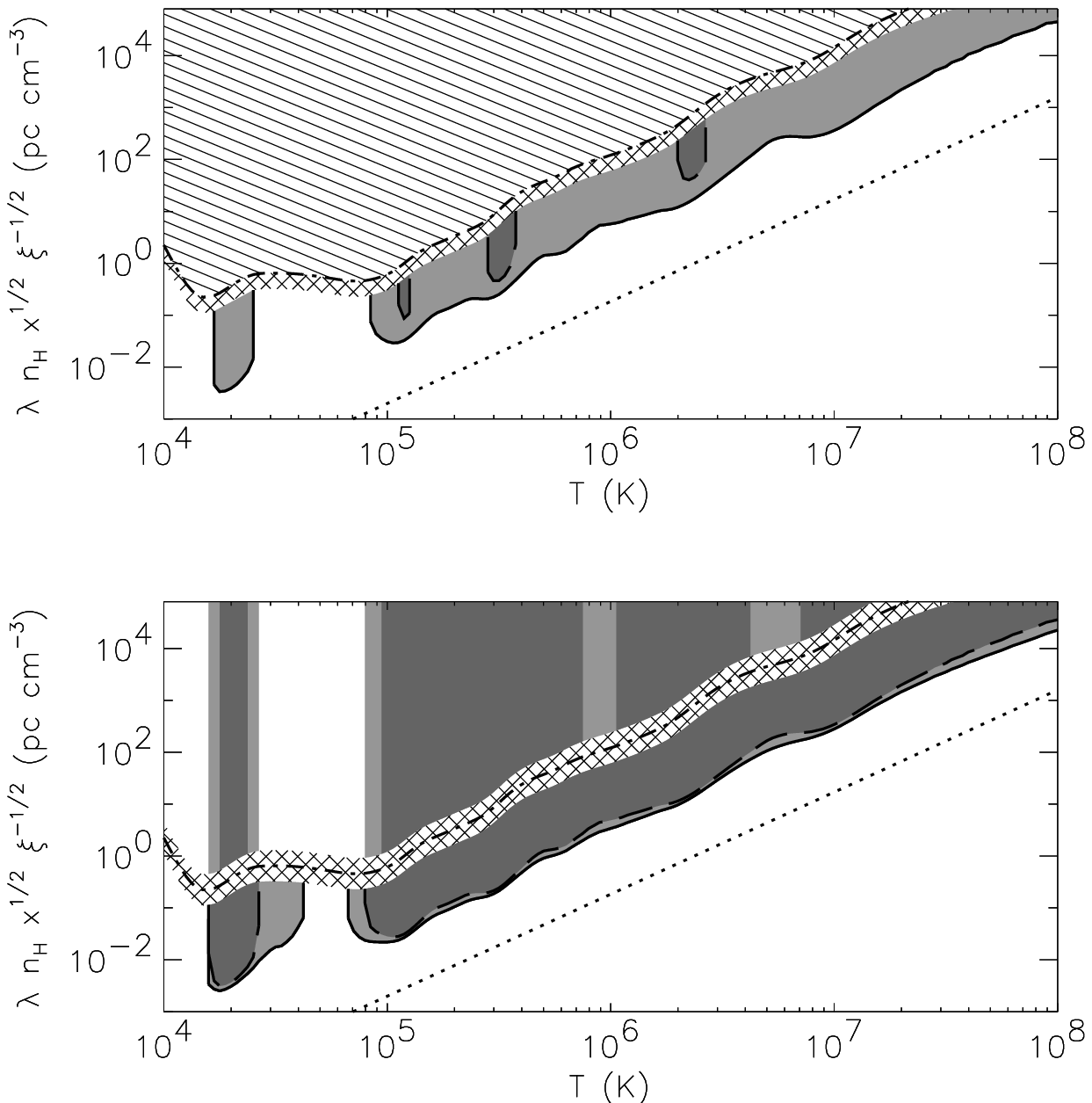}
\caption{Critical wavelengths and regions of instability for the sound waves
(upper panel) and the condensation mode (lower panel) for the application
discussed in section \ref{s_astr}.  The critical wavelength
$\lambcrit$ is plotted as a solid line for the case with net
background cooling and as a dashed line for the classical Field problem.
The dark-grey shaded area is the region where both problems yield
instability; in the light-grey shaded region only the case with net
background cooling is unstable. In the upper left corner of the upper panel
(hatched area in the figure) the sound waves are not properly defined so that no
statement concerning their stability is possible there. 
Two further lines depicted in the figure are the locus of points where
$\omega_s=\tau_\chi^{-1}$, with $\xi=1$ and $x=1$ (dotted) and
those where $\omega_s=\tau_c^{-1}$, with $\xi=1$ and
$x=1$ (dash-dotted line). The stability of the points in the figure in a
narrow band around the dash-dotted line (indicated with cross hatching) is
uncertain, as explained in the text.
\label{fig_apli3}}
\end{figure}

Figure \ref{fig_apli3} shows the instability regions and the critical lengths 
in the two--dimensional parameter space ($T$, $\lambda \, n_H \, x^{1/2}\,\xi^{-1/2}$).
 The upper diagram corresponds to the sound waves, the lower
one to the condensation mode. In the dark--grey shaded region, both the
classical Field problem and the case with net background cooling are
unstable, whereas in the light--grey shaded region only the latter is 
unstable. As expected, the instability region for the
sound waves is much larger for the case with net background cooling. For the
condensation mode, both regions are comparable, but, again, the range of
unstable temperatures is smaller for the classical Field problem.  In fact,
the ranges of instability for sound waves and condensation mode are similar
when there is net background cooling, which is in strong contrast to the
situation in the classical Field problem.  The upper left corner of each
diagram corresponds to a region where cooling is dominant (i.e., it occurs on
the fastest time-scale of the problem). Hence, there are no properly defined
sound waves in that region (hatched area in the upper diagram).

The critical length expression (\ref{eq_kcriapli}) is valid only when sound
is the dominant process, which happens within the region delimited by
the curves $\omega_s=\tau_\chi^{-1}$ (dotted line) and $\omega_s=\tau_c^{-1}$
(dash--dotted line) in the diagram. However, we know that the region below
the $\omega_s=\tau_\chi^{-1}$ -- line is stable, since conduction is
dominant. Similarly, in the cooling-dominated region (i.e., well above the
dash-dotted line), where conduction is negligible, we know the instability
properties of the condensation mode both for the classical Field problem and
for the case with net background cooling using an isochoric criterion as
explained in section \ref{s_cooling}. The remaining unknown region is, thus, a
(possibly narrow) band around the $\omega_s=\tau_c^{-1}$ -- line.
We have tentatively indicated this band in the figure as a cross--hatched
region. As a final remark, note that the use of the parameter $\lambda \, n_H \,
x^{1/2}\,\xi^{-1/2}$ in the figure is meaningful because we have ignored the
weak dependence of the Coulomb logarithm on $n_e$ (see
eq.~[\ref{eq_condspit}]). For the figure, $n_e$ has been set equal to
$1$cm$^{-3}$ in $\LambdaCoulomb$.

\subsection{Inclusion of a Background Magnetic Field without Dynamical Effects}\label{subsub_magn_field}

The astrophysical media considered in the previous subsection  (like the hot
interstellar medium or the intergalactic medium) are threaded by magnetic fields. Usually, these fields are
dynamically unimportant, that is, in these media $\beta>>1$, where $\beta$ is the ratio of
thermal and magnetic pressures. However, the effect of this weak magnetic
field on thermal conduction is very strong. Therefore, we include such a weak
magnetic field to complete the discussion of the previous paragraph. In a
medium threaded by a magnetic field, the growth rate of $\rho_1/\rho_0$ for
the condensation mode in the limit $\beta>>1$ is given by (A. J. Gomez-Pelaez
\& F. Moreno-Insertis, in preparation):
\begin{equation}\label{eq_condmag}
\Ss_{1c}=-\frac{(\gamma-1)}{\gamma}\frac{\mu}{\kboltz}\left\{ T_0
\left[\frac{\partial}{\partial T}
\left(\frac{\Ll}{T}\right) \right]_p +\frac{1}{\rho_0}\frac{k^2}{a^2} \left[{
\chi_\parallel}_0 \cos^2{\theta} + {\chi_\perp}_0 \sin^2{\theta}
\right] \right\}\;,
\end{equation} 
where $\theta$ is the angle between the background magnetic field,
$\bf{B_0}$, and the wavenumber vector $\bf{k}$, the conductivity
parallel to the magnetic field, $\chi_{\parallel}$, is given by equation 
(\ref{eq_condspit}), and the perpendicular conductivity, $\chi_{\perp}$, is given by \citep{spitzer}:
\begin{equation}\label{eq_cond_perp}
\frac{\chi_{\perp}}{\chi_{\parallel}}=7.2 \times 10^{-16} \left(\frac{\log
{\LambdaCoulomb}}{30} \right)^2 \frac{n_i^2}{B_{\mu G}^2 T_{6}^3}\;,
\end{equation}
in which $\log{\LambdaCoulomb}$ is given by equation (\ref{eq_condspit}), $n_i$ is the number
density of ions in cm$^{-3}$, $B_{\mu G}$ is the magnetic field in $\mu G$, and
$T_6$ is the temperature measured in $10^6$K$\,$. Expression
(\ref{eq_cond_perp}) is only applicable when the temperature of ions and electrons
is the same and the Larmor radius of the ions is much
smaller than their mean free path, which is always the case in the media considered in
this section. The corresponding expression for the sound waves is:
\begin{eqnarray}\label{eq_soundmag}
{\Ss_{1s}}_\pm^{} &=& -\frac{(\gamma-1)}{2}\frac{\mu}{\kboltz} \left\{
\frac{\gamma-1}{\gamma} \left[ \left( \frac{\partial \Ll}{\partial T}
\right)_S + \frac{1}{\rho_0}\frac{k^2}{a^2} \left( {\chi_\parallel}_0
\cos^2{\theta} + {\chi_\perp}_0 \sin^2{\theta} \right) \right] \right. 
\nonumber\\
\noalign{\vspace{4mm}}
  &-& \left. \left( \frac{3}{2}-\frac{1}{\gamma} \right) \frac{\Ll_0}{T_0}  
\right\} 
+\frac{(3\gamma -5)}{4}\frac{\dot{a}}{a} \;, 
\end{eqnarray}

In the limit $\beta>>1$, the only effect of the magnetic field is
to make the thermal conduction depend on the spatial direction. The growth rate
of both modes, and therefore their associated generalized Field lengths ($\lambcrit$), depend on
$\theta$. Setting both, $\Ss_{1c}$ (eq.~[\ref{eq_condmag}]) and ${\Ss_{1s}}_\pm^{}$
(eq.~[\ref{eq_soundmag}]), equal to zero, one obtains
\begin{equation}
\lambcrit(\theta) = \lambcrit_{\parallel} \left( \cos^2{\theta} +
\frac {\chi_\perp}{\chi_\parallel}\sin^2{\theta} \right)^{1/2},
\end{equation}
where $\lambcrit_{\parallel}$, is given by equation
(\ref{eq_kcriapli}). We will also use the symbol $\lambcrit_{\perp}$
for $\lambcrit(\theta=\pi/2)$.

\begin{figure} 
\plotone{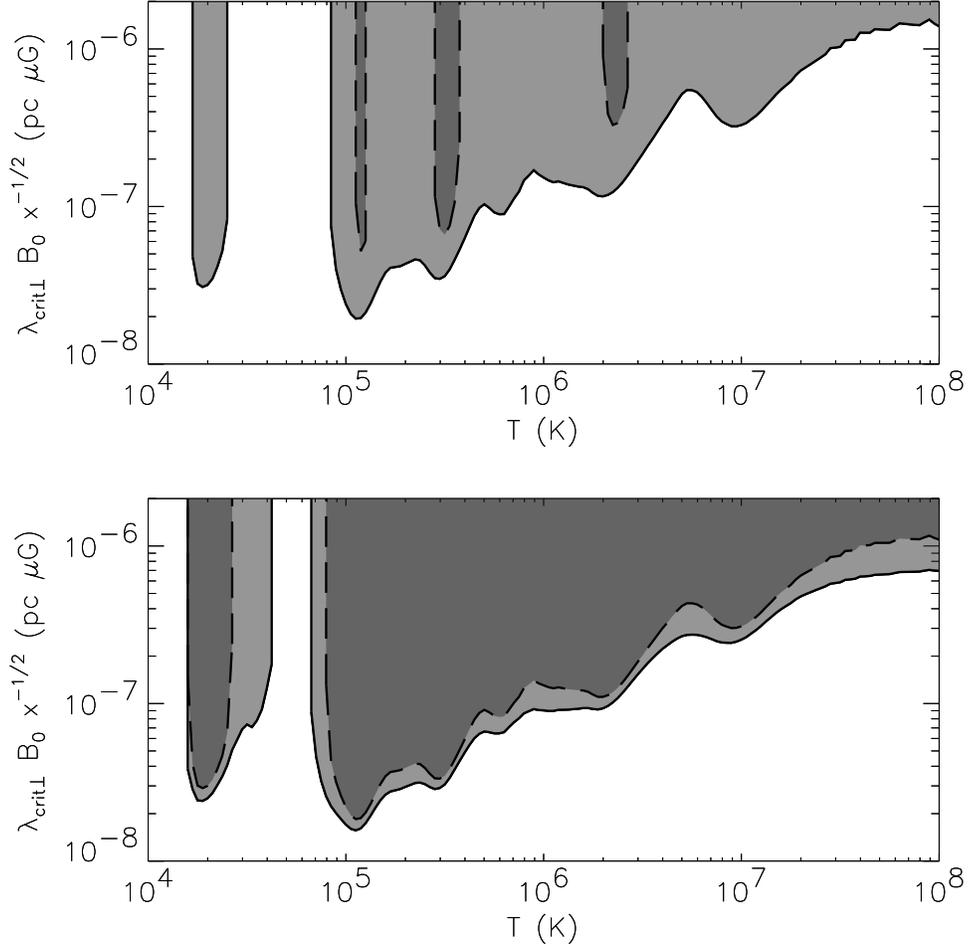}
\caption{Perpendicular Field length, $\lambcrit_\perp$, and regions of 
instability for a magnetized medium with $\beta \gg 1$
for the sound waves (upper panel) and the condensation mode (lower panel).
$\lambcrit_\perp$ is plotted as a solid line for the case with net background
cooling and as a dashed line for the classical Field problem.  In the dark-grey
shaded areas both the thermal equilibrium problem and the case with
net background cooling are unstable; in the light-grey shaded area only the
latter is unstable.
\label{fig_apli4}}
\end{figure}

Figure \ref{fig_apli4} shows the generalized (solid line) and classical
(dashed line) Field length and the instability regions for the condensation
and sound modes with $\theta=\pi/2$. The upper diagram corresponds to the
sound waves, the lower one to the condensation mode. Here again, dark-grey 
shading indicates that both the thermal equilibrium problem and the case 
with net background cooling are unstable; in the light-grey shaded area only 
the latter is unstable. Note that now the combination $\lambda \, B_{\mu G} 
\, x^{-1/2}$ appears in ordinates, since $\lambcrit_\perp$ does not depend 
on $n_H$ but on $B_{\mu G}$. From the figure (compare with
Figure~\ref{fig_apli3}), it is apparent that $\lambcrit_\perp <<
\lambcrit_\parallel$ and that $\lambcrit_\perp$ has a weaker dependence on 
temperature than $\lambcrit_\parallel$. Both features are a direct consequence of 
the conductivity ratio (\ref{eq_cond_perp}).  Hence, if there is a background
magnetic field, the condensations, in the linear stage, will be filaments
directed along the magnetic field lines with a length of the same order as
$\lambcrit_\parallel$ and a diameter of the same order as
$\lambcrit_\perp$.

Typical values of the critical length in the hot interstellar
medium (as inside SNRs and superbubbles) follow from the physical parameters
in them like: $n_H = 10^{-3}\;$cm$^{-3}$, $T = 10^6\;$K, and
$B_{\mu G} = 1$ \citep[e.g.,][]{mckee}. The associated critical lengths are:
(1) $\lambcrit_\parallel = 3.4\;$kpc and $\lambcrit_\perp = 9
\times 10^{-8}\;$pc for the condensation mode and (2)
$\lambcrit_\parallel = 6\;$kpc and $\lambcrit_\perp = 1.5 \times 10^{-7}\;$pc
for the sound waves. Note that $\lambcrit_\parallel$ is larger or of
the same order than the size of the considered systems whereas
$\lambcrit_\perp$ is well below it.

\section{SUMMARY AND DISCUSSION}
\label{ssum}

We have carried out a systematic analysis of the thermal stability of a
uniform medium which, in the unperturbed state, is undergoing net cooling
and expansion, so that the problem is {\bf not} amenable to a full Fourier
normal mode analysis. Thermal conduction and self-gravity have also been
included as fundamental ingredients. The small-perturbation problem yields a system of
ordinary differential equations with three independent solutions. In many
cases, two of them have the character of oscillatory modes (typically
sound waves, or, more generally, acoustic-gravity modes) and the third one
is a non-oscillatory (or condensation) solution. The wavelength of the
perturbation determines the positions of the different physical processes
in the hierarchy of timescales.  At small enough wavelength, thermal
conduction damps the condensation mode and makes sound waves isothermal.
At large wavelength, one or a few of the following processes are dominant: cooling,
self-gravity and expansion.  For example, if cooling is the only dominant
process, the perturbation evolving on the fastest timescale is isochoric
with no sound waves present that could even out the pressure gradients. At
intermediate wavelength, the propagation of sound is the fastest process,
so the evolution of the condensation mode is isobaric.  In the paper,
solutions have been obtained in a number of instances using the WKB
method. WKB solutions can  be found only if two conditions are fulfilled: 
(1) there is a clear separation of timescales between {\it fast} and {\it
slow} physical processes in the system, and (2) the background expansion and 
cooling belong to the {\it slow} category.

Of special interest in this paper are those solutions whose growth rate is
of the same order as the rate of change of the unperturbed background. In
this case, the change in time of the critical wavelengths (the generalized
Field length, $\lambcrit$, and the Jeans length, $L_J$) may make the
perturbation go through totally different stability regimes during their
evolution. For example, the solutions obtained in section \ref{s_sougr} 
for a medium undergoing net cooling include a condensation mode initially 
in the stable range whose wavelength becomes larger than the Field length 
after some time, thus becoming thermally unstable. Later on, the Jeans 
length decreases sufficiently so that the perturbation also becomes 
gravitationally unstable. A sound wave may undergo a similar process of change of
stability, first being in a stable regime with slowly decreasing
amplitude, then becoming thermally unstable and finally, when the
self-gravity frequency becomes comparable with the sound frequency, turning 
into an acoustic-gravity mode which ends up as a gravitationally-dominated 
perturbation.

The analysis of the present
paper shows (section \ref{s_astr}) that in a medium undergoing a
realistic net cooling the sound waves are thermally unstable for a
comparable range of wavelengths and temperatures as the condensation mode. 
This may come as a surprise: in contrast, and by
way of example, if the thermal equilibrium is reached by balancing the
cooling with a constant heating, the sound waves are stable for
almost all temperatures in the range $10^4$ -- $10^8$ whereas the condensation
mode continues being unstable. On the other hand, the inclusion of a weak
magnetic field has a strong influence on the thermal conduction. For
astrophysical environments such as the hot ISM and the hot IGM, the
generalized Field length along the magnetic field is many orders of
magnitude larger than in the perpendicular direction. Therefore, the
unstable condensations will be filaments directed along the magnetic field
lines. 

For the description of the astrophysical systems in which thermal
instabilities appear, it is important to study the non-linear phase of
both condensation mode and sound waves. The unstable sound waves develop
pairs of shock fronts and rarefaction waves, as known from elementary
hydrodynamics. The condensation mode, instead, just continues growing
without propagating. The detailed physics of the nonlinear phase has to be
calculated, in general, using numerical means. Yet, a number of results can
be advanced at this stage. For instance, in the linear analysis, a
perturbation with negative $\rho_1/\rho_0$ grows (in absolute value) as
fast as a perturbation with positive $\rho_1/\rho_0$. This symmetry is
broken in the non-linear evolution. In a medium undergoing net cooling,
for example, a condensation with positive $\rho_1/\rho_0$ (negative
$T_1/T_0$) grows faster than one with $\rho_1/\rho_0 < 0$. This happens
because of the sum of the following two processes. First, thermal
conduction grows with temperature. Therefore, a condensation with positive
$T_1/T_0$ is more damped by thermal conduction than a negative one.
Second, in the range $10^5 - 10^7$ cooling increases toward lower
temperatures (Fig.~\ref{fig_apl1}). Therefore, a condensation with
negative $T_1/T_0$ is more destabilized by cooling than a positive one. A
related aspect is that the characteristic spatial size (width at half
height) of a condensation decreases by orders of magnitude during its
evolution. This happens because the coldest zones of the perturbation cool
much faster than the warmer ones. The last aspect can be seen in the paper
of \citet{david88}, where the non-linear evolution of a perturbation of
this type is computed numerically. 

In the weakly--magnetized case, at the
beginning of the non-linear evolution ($\beta \gg 1$), the compression in the
filament proceeds mainly in the directions perpendicular to the magnetic field
lines. This is because of the reduction of thermal conduction transversely
to the field lines: a much higher temperature (and thus pressure) gradient
can be maintained across than along the field lines. The plasma $\beta$
inside the condensation decreases with time because of the temperature
decrease and the growth of magnetic field due to the compression. In the
advanced non-linear phase, the perpendicular compression comes to a halt
once $\beta \ll 1$, since then the net force across the field lines (the
gradient of the sum of magnetic and thermal pressures) is almost decoupled
from temperature. After this, the compression is only possible along the
field lines. As a related example, see \citet{david89}, which numerically
compute the non-linear evolution of a planar condensation taking into
account magnetic pressure.
 
The WKB analysis that we have carried out in this paper assumes a uniform
background. However, our results are also applicable for a weakly
non-uniform background. More in detail, our results for the condensation
mode in the sound domain are valid if the Brunt-V\"ais\"al\"a frequency
associated with the background entropy gradient is much smaller than the
growth rate of the condensation mode. This holds, for example, in the
supersonic region of a wind and inside old SNRs and superbubbles. In
contrast, for the study of the thermal instability in cooling flows in
clusters of galaxies, the buoyancy of the perturbations must be taken into
account. Many authors have studied this problem: \citet{mathews},
\citet{malagoli}, \citet{balbussoker89}, \citet{loew} and
\citet{balbus91}.  \citet{loew} and \citet{balbus91}, in particular,
conclude that a magnetic field, even a weak one, can inhibit buoyancy and
trigger thermal instability, which otherwise would be small due to
buoyancy. The extension of the results of the present paper to
environments of this type requires additional work.

\acknowledgments

Financiation through DGES project PB95-0028C of the Spanish Ministry of
Education is acknowledged. The authors are grateful to Dr J.~C.~Raymond for
providing them with the cooling functions used in this paper.

\appendix
 
\section{COEFFICIENTS FOR THE WKB SOLUTION}

\subsection{Coefficients for the Sound Domain}
\label{ApCa}

For the sound domain case (see sec. \ref{ssou}), the different order
components of the coefficients $A(t)$, $B(t)$, and $C(t)$, have the value:
\begin{equation}\label{coesd1}
A_3=0 \;, \quad B_2=\omega_s^2 \;, \quad  C_1=0 \;,
\end{equation}
\begin{equation}\label{coesd2}
A_2=\frac{\omega_s^2}{\gamma} \left(\Omcdp + \tchi^{-1}\right) \;, \quad B_1=0 \;, \quad
C_0= \Omc + \Omcdrho + \tchi^{-1} + (3\gamma +1)\frac{\dot a}{a} \;,
\end{equation}
\begin{equation}\label{coesd3}
A_1=0\;, \quad A_0= \omega_g^2\left[(4-3\gamma) \frac{\dot a}{a}-\Omc - \Omcdrho - \tchi^{-1}  \right]\;,
\end{equation}
\begin{equation}\label{coesd31}
B_0 = 2\frac{\ddot{a}}{a} -\omega_g^2 - 2\frac{\dot a}{a} \left[(2-3\gamma)
\frac{\dot a}{a}- \Omc  - \Omcdrho - \tchi^{-1} \right]\;.
\end{equation}

\subsection{Coefficients for the Sound and Self-Gravity Domain without Expansion}
\label{ApE}

When sound and self-gravity dominate and there is no expansion (see sec. \ref{s_sougr}),
the different order components of the coefficients $A(t)$, $B(t)$, and $C(t)$, have the value:
\begin{equation}\label{coesgd1}
A_3=0 \;, \quad B_2=\omega_s^2-\omega_g^2 \;, \quad  C_1=0 \;,
\end{equation}
\begin{equation}\label{coesgd2}
A_2=\frac{\omega_s^2}{\gamma}\left(\Omcdp + \tchi^{-1}\right) -\omega_g^2 \left[\Omc+ \Omcdrho + \tchi^{-1}  \right]\;,
\quad B_1=0 \;, \quad C_0=\Omc + \Omcdrho + \tchi^{-1}\;.
\end{equation}

\end{document}